\documentclass[aps,pra,showpacs,superscriptaddress, twocolumn,final,numerical,10pt]{revtex4-1}

\usepackage{epsfig}
\usepackage{graphicx}
\usepackage{dcolumn}
\usepackage{bm}
\usepackage{amsthm}
\usepackage{amsfonts}
\usepackage{amscd}
\usepackage{amsmath}
\usepackage{amssymb}
\usepackage{enumerate}
\usepackage{color}
\usepackage{lipsum}

\usepackage{mathcomp} 
\usepackage{ulem} 

\newcommand{\ket}[1]{\mbox{$ | #1 \rangle $}}
\newcommand{\bra}[1]{\mbox{$ \langle #1 | $}}

\newcommand{\beq}{\begin{equation}}
\newcommand{\eeq}{\end{equation}}
\newcommand{\bea}{\begin{eqnarray}}
\newcommand{\eea}{\end{eqnarray}}
\def \( {\left(}
\def \) {\right)}
\def \[ {\left[}
\def \] {\right]}

\begin{document}

\title{Information theoretic security of quantum key distribution overcoming the
repeaterless secret key capacity bound}

\author{Kiyoshi Tamaki}
\address{Graduate School of Science and Engineering for Education, University of Toyama, Gofuku 3190, Toyama 930-8555, Japan }
\author{Hoi-Kwong Lo}
\author{Wenyuan Wang}
\address{Center for Quantum Information and Quantum Control, Department of Physics and Dept. of Electrical \& Computer Engineering, University of Toronto, M5S 3G4 Toronto, Canada}
\author{Marco Lucamarini}
\address{Toshiba Research Europe Ltd, 208 Cambridge Science Park, Cambridge CB4 0GZ, United Kingdom}

\begin{abstract}
\noindent
Quantum key distribution is a
way to distribute secret keys to distant users with information theoretic security and key rates suitable for real-world applications. Its rate-distance figure, however, is limited by the natural loss of the communication channel and can never surpass a theoretical limit known as point-to-point secret key capacity.
Recently, a new type of quantum key distribution with an intermediate relay was proposed to overcome this limit (M. Lucamarini, Z. L. Yuan, J. F. Dynes and A. J. Shields, Nature, 2018). 
However, a standard application of the decoy state method limited the security analysis of this scheme to hold under restrictive assumptions for the eavesdropper.
Hence, overcoming the point-to-point secret key capacity with an information-theoretic secure scheme is still an open question.
Here, we propose a novel way to use decoy states to answer this question. The key idea is to
switch between a Test mode and a Code mode, the former enabling the decoy state parameter estimation and the latter
generating a key through a phase encoding protocol. This way, we  confirm the scaling properties of the original scheme and overcome the secret key capacity at long distances.
Our work plays a key role to unlock the potential of practical secure quantum communications.
\end{abstract}

\maketitle

\section{Introduction}
\label{sec:intro}

\noindent Quantum key distribution (QKD)~\cite{BB14} makes it possible to distribute cryptographic keys to remote users with security that is independent of an attacker's computational power~\cite{review}, a feature denoted `information-theoretic security'. After several years of development, QKD is now gaining momentum and is being deployed worldwide, mainly in the form of quantum networks~\cite{wiki-quantum}. To maintain and reinforce this positive trend, it is important to look at practical applications and tackle the problems that currently limit this technology.

An often-mentioned obstacle in QKD is the circumscribed maximum distance at which keys can be distributed. Despite the intensive research on quantum repeaters~\cite{BDC+98,DLC+01,SSD+11} and on relaxing their technological demands~\cite{JTN+09,MSD+12,LJL16, ATL15,ATM15}, there are no cheap and efficient solutions to repeat an unknown quantum signal along the transmission line yet, in a fashion similar to a repeater in standard optical communications. Without a quantum repeater, the QKD signal unavoidably faces exponential loss during the propagation in the optical medium and becomes too small to be faithfully measured by the noisy detectors at the receiving side.

Even with noiseless detectors, it is impossible, in fact, to increase rate and distance of QKD beyond a certain limit, a result recently proven in~\cite{TGW14,PLO+17, N16}. The point-to-point secret key capacity of a quantum channel~\cite{PLO+17}, which we denote simply ``SKC",  upper bounds the maximum secret information that can be transmitted via QKD on an uninterrupted link characterised only by its transmission $\eta$~\cite{TGW14}, irrespective of the amount of noise it presents. To overcome the SKC, quantum repeaters were believed to be necessary.

Recently, however, it was shown that it is possible to overcome the SKC without using quantum repeaters, with a scheme named ``Twin-Field QKD'' (TF-QKD)~\cite{LYD+18} built only with presently available components and an intermediate relay. TF-QKD is similar to the decoy-state~\cite{decoy} measurement-device-independent (MDI) QKD~\cite{mdiQKD}, but it allows for a much higher rate-distance figure, as it is based on single-photon detections rather than on two-photon detections. Quite remarkably, all the other positive features of MDI-QKD, like its tolerance to detectors vulnerabilities and its readiness for star networks~\cite{RLY+17}, are retained by TF-QKD.

On the other hand, TF-QKD does not offer yet the information-theoretic security demanded by QKD. In fact, proving the full security of TF-QKD was left as an important open question in the original paper~\cite{LYD+18}. Answering this question should clarify whether it is possible to overcome the SKC with an information-theoretic secure scheme.

The difficulty encountered in~\cite{LYD+18} is related to the use of decoy states~\cite{decoy}, which are key to long-haul quantum communications. To enable decoy states, the phase of the states initially prepared by the users (Alice and Bob) should be random and unknown to the eavesdropper (Eve). Therefore the random phases are usually kept secret and this guarantees that Eve can only see a mixture of photon number states. In TF-QKD, however, the random phases are revealed, to allow Alice and Bob reconcile their data, and this could help Eve in her attacking strategy.

This possibility has been ruled out in~\cite{LYD+18} by restricting Eve's attack to those that commute with the photon number operator. However this prevents the information-theoretic security of TF-QKD, which demands no assumptions
on Eve. 
It is then essential 
to show the security of this scheme with a rigorous security proof.

Here, we introduce a TF-QKD protocol, which we call TF-QKD*, that overcomes the SKC limit and at the same time is information-theoretic secure.
Our key idea is to select between a Test mode and a Code mode probabilistically, allowing us
to use decoy states and the phase of the states. Similarly to TF-QKD, this protocol's key rate scales with the square-root of the channel transmission, $\sqrt{\eta}$, thus entailing a major improvement in the tolerance of the channel loss.
Importantly, because this protocol's key rate represents a lower bound valid against any attacks
allowed by the laws of physics, we rigorously prove that it is possible to surpass the SKC without using quantum repeaters, as conjectured in~\cite{LYD+18}.

\section{TF-QKD* protocol}
\label{sec:II}

In this section we introduce our key idea, which is to distinguish between a Test mode, in which the phases are not disclosed by the users and the decoy-state method~\cite{decoy} is applied, and a Code mode, where the phases are disclosed and a key is generated. We start from the description of the protocol, which is given below. In it we assume that the random phases $\theta_{\textrm{A}}$ and $\theta_{\textrm{B}}$ are automatically generated by the users with uniform distribution. So we skip the step for generating these phases in the protocol.


\begin{figure}
\begin{center}
 \includegraphics[width=1\columnwidth]{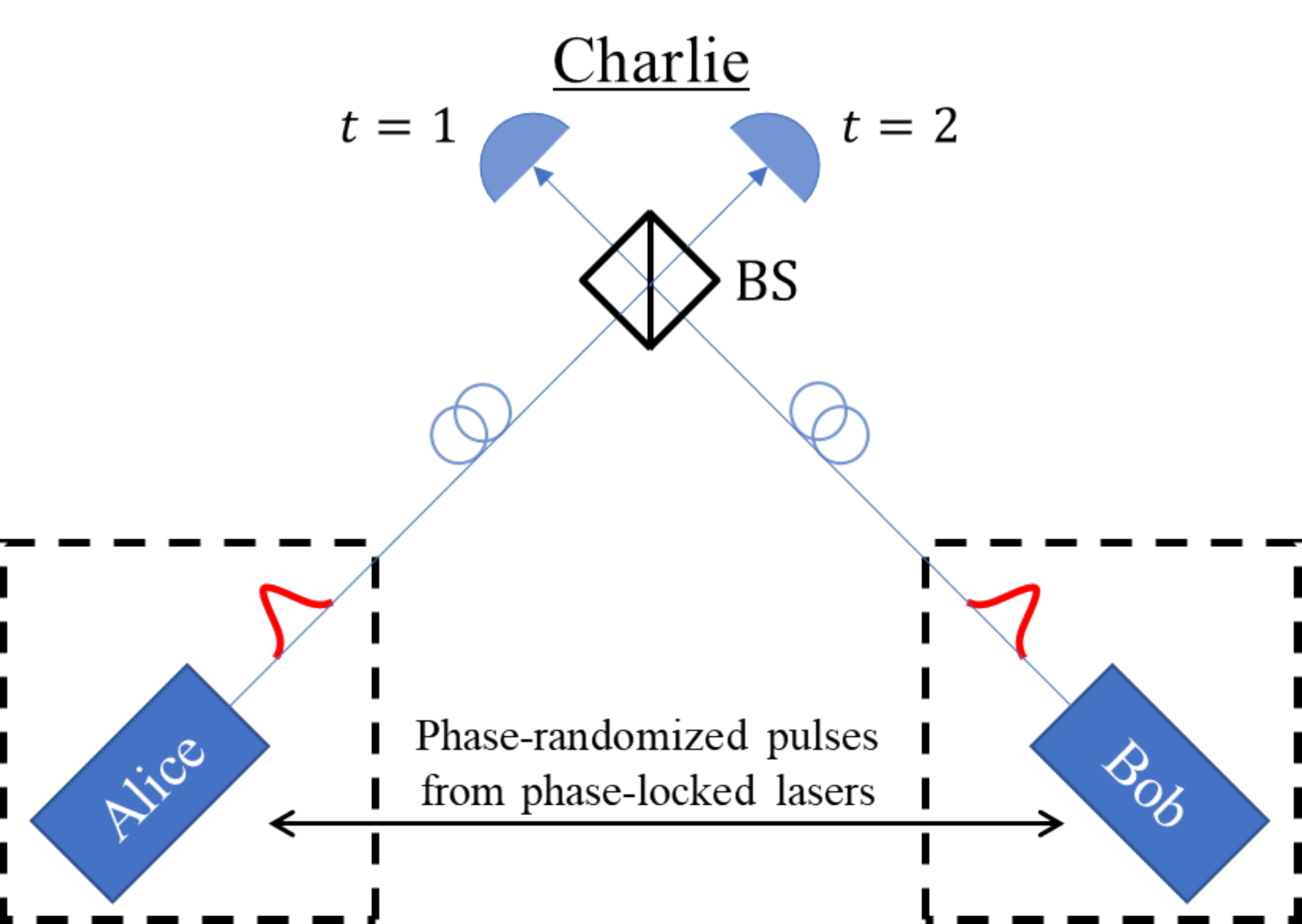}
 \end{center}
 \caption{Schematics for the experimental setup for TF-QKD*. The scheme is essentially the same as the one for the original TF-QKD. 
Here, Alice and Bob's lasers are phase locked, and before sending a pulse, each
of Alice and Bob's pulse is phase-randomized independently, and BS represents a 50:50 beam splitter. If the detector corresponding to $t=1$ ($t=2$) clicks, Charlie is supposed to announce  $t_{\rm E}=1$ ($t_{\rm E}=2$), and if both of the detectors click, then he is supposed to choose $t_{\rm E}=1$ or $t_{\rm E}=2$ at random and announces his choice.}
 \label{experimental setup}
\end{figure}

\vspace{0.7cm}

{\small

{\bf \noindent TF-QKD* protocol}
\begin{enumerate}
\item Alice and Bob repeat Steps 2-3, $N$ times. All the public announcements by Alice and Bob are done over an authenticated channel.

\item Alice (Bob) randomly chooses a bit value $j_{\rm A}\in\{0, 1\}$ ($j_{\rm B}\in\{0, 1\}$). Next, Alice (Bob) chooses
the following quantities with the following probabilities:

Basis: $b_{\rm A}\in\{{\rm Z_A}, {\rm Y_A}\}$ ($b_{\rm B}\in\{{\rm Z_B}, {\rm Y_B}\}$) with probability $p_{\rm Z_A}$ and $p_{\rm Y_A}$ ($p_{\rm Z_B}$ and $p_{\rm Y_B}$), respectively. For simplicity we set $p_{\rm Z_A}=p_{\rm Z_B}$ and $p_{\rm Y_A}=p_{\rm Y_B}$.

Intensity: $\mu_{\rm A} \in \{\mu_1, \mu_2, \mu_3\}$ ($\mu_{\rm B} \in \{\mu_1, \mu_2, \mu_3\}$) with probability $p_{\mu_{1}}$, $p_{\mu_{2}}$, and $p_{\mu_3}$ ($p_{\mu_{1}}$, $p_{\mu_{2}}$, and $p_{\mu_3}$), respectively.
 
Then, Alice (Bob) prepares a coherent signal $(\ket{e^{i (\theta_{\rm A}+
\delta_{\rm A})}\sqrt{\mu_{\rm A}}}_{\rm sg})_{\rm E1}$ ($(\ket{e^{i (\theta_{\rm B}+
\delta_{\rm B})}\sqrt{\mu_{\rm B}}}_{\rm sg})_{\rm E2}$), where $\delta_{\rm A}=j_{\rm A}\pi$ 
($\delta_{\rm B}=j_{\rm B}\pi$) for $b_{\rm A}={\rm Z_A}$ ($b_{\rm B}={\rm Z_B}$) and
$\delta_{\rm A}=3\pi/2-j_{\rm A}\pi$ ($\delta_{\rm B}=3\pi/2-j_{\rm B}\pi$) for $b_{\rm A}={\rm Y_A}$
($b_{\rm B}={\rm Y_B}$). Here,
the subscript `sg' is added to emphasize that this is a signal pulse to be transmitted.

Finally, Alice (Bob) measures the phase information $\theta_{\rm A}$ ($\theta_{\rm B}$) and sends system E1 (E2) over the quantum channel.

\item
Charlie measures the incoming signals. Ideally, he is supposed to perform a single photon counting measurement on systems sg in E1 and E2, and if he obtains a double click, then
he randomly decides $1$ or $2$ (see Fig. \ref{experimental setup}). However, he could do anything he pleases.

If Charlie obtains a detection event, he announces this as well as the type of the outcome $t_{\rm E}\in\{1, 2\}$
over a public channel. If Charlie announces outcomes other than $1$ or $2$, including a non-detection outcome, Alice and Bob discard all the data associated with this event.

\item
For each of the detection events, Alice and Bob announce their intensity selections. Also, depending on whether $\mu_{\rm A}=\mu_{\rm B}$ is satisfied or not, they conduct the following operations:

(i) If $\mu_{\rm A}\neq \mu_{\rm B}$, they announce their basis selections.

(ii) If $\mu_{\rm A}=\mu_{\rm B}$, Alice randomly assigns each instance to the Test mode or the Code mode, with probabilities $p_{\rm T}$ or $p_{\rm C}$, respectively, and announces her choice.

(ii-i) If the Test mode was selected, then Alice and Bob announce their bases.

(ii-ii) If the Code mode was selected, Alice (Bob) announces the phase information $\theta_{\rm A}$ ($\theta_{\rm B}$).
Alice then chooses one of two bases, ${\rm Z_C}$ or ${\rm X_C}$, with probabilities $p_{\rm Z_C}$ and $p_{\rm X_C}=1-p_{\rm Z_C}$, respectively, and announces her selection.

When the ${\rm Z_C}$ basis was selected, Alice and Bob announce the basis information
$b_{\rm A}$ and $b_{\rm B}$ that were selected in Step 1. If $b_{\rm A}\neq b_{\rm B}$, Alice and Bob discard
the instance, and if $b_{\rm A}= b_{\rm B}$, they
announce the bit values $j_{\rm A}$ and $j_{\rm B}$ except for ${\rm Z_A}={\rm Z_B}$. When the ${\rm X_C}$ basis was selected, Alice and Bob announce nothing.

Finally, when $|\theta_{\rm A}-\theta_{\rm B}|\le\Delta/2$ ($|\theta_{\rm A}-\theta_{\rm B}|>\Delta/2$), Alice and Bob keep (discard)
the corresponding outcomes. However, they keep the record of the number of the outcomes occurred even if they discard the data.

\item For each of the bit string with $\mu_{\rm A}=\mu_{\rm B}=\mu$, if it originates from $t_{\rm E}=2$ Alice flips
her bit string, and if it originates from $t_{\rm E}=1$ Alice does nothing on her bit string.
Then, for each of the bit string with $\mu_{\rm A}=\mu_{\rm B}=\mu$ and $t_{\rm E}$, Alice and Bob apply
error correction by exchanging a syndrome information encrypted by a previously shared secret key. Then, Alice
selects a hash function randomly according to the result of a parameter estimation
based on the data in Step 4, 
and announces the selected function. Alice and Bob perform privacy amplification based on 
the selected Hash function to share a key.
\end{enumerate}
}

\noindent There are some remarks about this actual protocol:
\begin{itemize}
  \item[$\diamond$] In the Test mode, Alice and Bob do not announce the phase information measured in Step 2. Rather they keep
it secret
from Eve. This way, the state that each of Alice and Bob sends in the Test mode can be regarded as classical mixtures of
number states from Eve's viewpoint, enabling Alice and Bob to employ the decoy state method in the Test mode.
  \item[$\diamond$] Although we made a redundant definitions of $p_{\rm Z_A}$, $p_{\rm Z_B}$, $p_{\rm Y_A}$, and $p_{\rm Y_B}$ such that
$p_{\rm Z_A}=p_{\rm Z_B}$ and $p_{\rm Y_A}=p_{\rm Y_B}$ hold, we explicitly use these different variables to denote Alice's probability and Bob's probability for clarity of the discussions.
  \item[$\diamond$] TF-QKD* protocol generates a key separately depending on $\mu_{\rm A}=\mu_{\rm B}=\mu$ and $t_{\rm E}$.
  \item[$\diamond$] In Step 4 (iii), the presence of the choice between ${\rm Z_C}$ and ${\rm X_C}$ entails a loss of the generated key unless ${\rm Z_C}$ is chosen. Also, $p_{\rm Z_C}$ is a parameter that has to be optimized in the finite key regime.
  \item[$\diamond$] When the ${\rm X_C}$ basis is chosen in the Code mode, Alice and Bob do not announce their bases. This means that Alice and Bob do not know whether their bases selections made in Step 2 coincide or not. However, the number of this events can be estimated
from the event with the ${\rm Z_C}$ basis in the Code mode.
\end{itemize}
Let us now provide an intuitive picture of this protocol and the main reason why it is secure. Suppose that Alice transmits a phase-randomized coherent pulse over a quantum channel, and Alice keeps the phase information in her lab. From Charlie's viewpoint, her state can be described as
\begin{eqnarray}
\int_{0}^{2\pi} d\theta \ket{e^{i\theta}\sqrt{\mu}}_{\rm E1}\bra{e^{i\theta}\sqrt{\mu}}\,.
\label{density operator}
\end{eqnarray}
Here, the subscript ${\rm E1}$ refers to the system of the pulse, and $\mu$ is the mean photon number of the pulse.
In the decoy state method, the fact that Alice's state can be regarded as a classical mixture of number states is fundamental. On the other hand, the phase of Alice's state plays a key role in a protocol where the phase of the encoded pulses is used to generate a key.
Therefore, we discuss two observables of the system ${\rm E1}$, the photon number and the phase, and by recalling
that any observable is expressed through measurements in quantum physics, it is convenient to
introduce another system ${\rm P_A}$ which purifies the states as follows
\begin{eqnarray}
\int_{0}^{2\pi} \frac{d\theta}{\sqrt{2\pi}}   \ket{\theta}_{\rm P_A}\ket{e^{i\theta}\sqrt{\mu}}_{\rm E1}&=&e^{-\frac{\mu}{2}}\sum_{n=0}^{\infty} \frac{\sqrt{\mu}^{n}}{\sqrt{n!}}\ket{n}_{\rm P_A} \ket{n}_{\rm E1}\nonumber\\
&=& \sum_{n=0}^{\infty} \ket{n}_{\rm P_A} {\hat P}_{n}^{({\rm E1})}\ket{e^{i\theta}\sqrt{\mu}}_{\rm E1}\nonumber\\
\label{identity}
\end{eqnarray}
where
\begin{eqnarray}
\ket{n}_{\rm P_A}=\int_{0}^{2\pi} \frac{d\theta}{\sqrt{2\pi}} e^{in\theta}\ket{\theta}_{\rm P_A}\,,
\label{number state}
\end{eqnarray}
and ${\hat P}_{n}^{({\rm E1})}$ is a projection operator to a $n$ photon space of system ${\rm E1}$.
Here, $\theta$ is defined within the interval $[0, 2\pi)$, $\bra{\theta'}\theta\rangle=\delta (\theta'-\theta)$ with $\delta(x)$ being the Dirac's delta function, and one can show the standard relationship ${}_{\rm P}\langle m\ket{n}_{\rm P}=\delta_{m,n}$ (see Appendix~\ref{phase-number-relationship}
for details).

From these equations, it is clear that when Alice obtains the photon number information of system ${\rm P_A}$, then the information about the phase is destroyed. A direct consequence of this is that Alice cannot employ the decoy state method when she measures the phase. In other terms, the two observables corresponding to global phase and photon number of the same quantum system do \textit{not} commute~\cite{phase-number}.

Our key idea to overcome this problem is to introduce a Test mode and a Code mode in the protocol, which
are probabilistically chosen by
Alice after the transmission of pulses. When the Code mode is chosen, Alice announces the phase information, and the users employ
a phase encoding scheme similar to phase-based MDI-QKD~\cite{mdiQKD, TLF+12}
to generate a key, but without resorting to the decoy state method. On the other hand, when the Test mode is selected, Alice does not announce the phase information, so the users
can employ the decoy state method to estimate the parameters needed for the security of the phase-based MDI-QKD in the Code mode.

In more detail, 
the main parameter to be estimated in the Code mode is the bias of an $X$ basis measurement on a fictitious system called the ``quantum coin''.
This bias is a key parameter to represent a basis dependency of the pulses arising from the non-randomized phase~\cite{GLLP, LP06, koashi09}. The smaller the bias, the better the key rate we can achieve.

In the literature~\cite{TLF+12, LP06}, it is simply assumed the worst case scenario, where Charlie enhances the bias of the quantum coin by exploiting the channel loss. 
This results in a dramatic reduction of the key generation rate. In contrast, with our idea, the decoy state method in the Test mode
provides a tight estimation of the bias in the Code mode and we do not have to rely
on the worst case scenario, leading to a significant improvement in key generation rate. This tight estimation is possible because Alice chooses
between the Code and Test modes after Charlie announces his measurement outcome, and as a result, Charlie or Eve cannot behave
differently between the two modes. Therefore, the bias in the Test mode serves as a good
sample of that in the Code mode, and we can use the random sampling theory to estimate
the bias in the Code mode from the one in the Test mode (see Appendix~\ref{Appendix-securityproof-asymt} for more detail). In the proof, we consider that this bias is enhanced due to the post-selection depending on whether $|\theta_{\rm A}-\theta_{\rm B}|\le\Delta/2$ or not. Here, importantly, unlike the worst case scenario in~\cite{TLF+12, LP06}, 
this enhancement is not dependent on the channel losses, but it depends only on 
a constant factor, $2\pi/\Delta$. Therefore, this enhancement does not affect the key rate drastically.

The security proof of TF-QKD* protocol is given in the Appendix \ref{Appendix-securityproof-asymt}. There, we assume the use of infinite number of decoy states in the limit of large key size, for simplicity. However, in Appendix~\ref{complete proof}, we provide the complete information theoretic security proof using three decoy states in the finite key size. In the following section, we present the result of a simulation for this protocol in the asymptotic case, for simplicity. Then, we conclude with the final remarks in Sec.~\ref{main-conclusion}.

\section{Simulation of the key rate}
In this section, we simulate the key generation rate based on our security proof. For simplicity, we assume that the number of the decoy
states is infinite and the number of pulses sent is large enough to neglect any statistical fluctuation,
and furthermore, we consider that Alice and Bob choose the ${\rm Z}$ and ${\rm Y}$ bases with the same
probability, i.e. $p_{\rm Z_A}=p_{\rm Z_B}=p_{\rm Y_A}=p_{\rm Y_B}$.
In this case, when Alice and Bob select the same intensity setting $\mu$ in the Code mode and Charlie announces $t_{\rm E}$, we can write the key rate as~\cite{key length, SCIC}
\begin{eqnarray}
l_{\mu, t_{\rm E}}&=&N_{{\rm sif}, {\rm Z}, \mu, t_{\rm E}}\left[1-h\left(e_{{\rm ph}, \mu, t_{\rm E}}\right)\right]-\lambda_{{\rm EC}}\,,
\label{rough key rate}
\end{eqnarray}
where $N_{{\rm sif}, {\rm Z}, \mu, t_{\rm E}}$ is the length of the ${\rm Z}$ basis sifted key, i.e. a bit string in
which Alice and Bob agree with the ${\rm Z}$ basis in the Code mode, they use a particular intensity choice $\mu$, Charlie announces $t_{\rm E}$, and
$|\theta_{\rm A}-\theta_{\rm B}|\le\Delta/2$.
$h(x):=-x\log_{2}x-(1-x)\log_{2}(1-x)$ is the binary entropy and $\lambda_{{\rm EC}}$ is the amount of information
exchanged for error correction. An important parameter in Eq. (\ref{rough key rate}) is the phase error rate
$e_{{\rm ph}, \mu, t_{\rm E}}$, which is related to
the amount of privacy amplification and is expressed by \cite{LP06}
\begin{eqnarray}
e_{{\rm ph}, \mu, t_{\rm E}}&=& e_{Y_{\rm er}, \mu, t_{\rm E}}+4\Delta_{\rm bias}(1-\Delta_{\rm bias})(1-2e_{Y_{\rm er}, \mu, t_{\rm E}})\nonumber\\
&+&
4(1-2\Delta_{\rm bias})\sqrt{\Delta_{\rm bias}(1-\Delta_{\rm bias})e_{Y_{\rm er}, \mu, t_{\rm E}}}\nonumber\\
&\times&\sqrt{(1-e_{Y_{\rm er}, \mu, t_{\rm E}})}\,.
\label{main-simplified-ineq-phase}
\end{eqnarray}
Here, $e_{Y_{\rm er}, \mu, t_{\rm E}}$ is an error rate in the the ${\rm Y}$ basis sifted key, and $\Delta_{\rm bias}$ is the
bias we estimate from the Test mode by exploiting the decoy state method.

As a channel model, we suppose that
bit errors stem from the dark count and/or the intrinsic bit errors of TF-QKD${}^{*}$ due to the phase difference
$|\theta_{\rm A}-\theta_{\rm B}|\le\Delta/2$, and we neglect system errors, such as misalignment errors.
In the simulation, we assume that $\Delta=2\pi/8$, and the transmission rate of a quantum channel
is represented by $e^{-\alpha L/10}$ with $\alpha=0.2$ and $L$ the length of an optical fibre.
Moreover, we assume the efficiency of the error correcting code is $1.1$. As for the detectors used by Charlie, we assume rather a practical parameters~\cite{scontel} for detection efficiency $\eta_{\rm det}=80\%$, and we assume dark count rate $p_{\rm dark}=1.0\times 10^{-11}$ per pulse~\cite{NTT-det}.

\begin{figure}
\begin{center}
 \includegraphics[width=1\columnwidth]{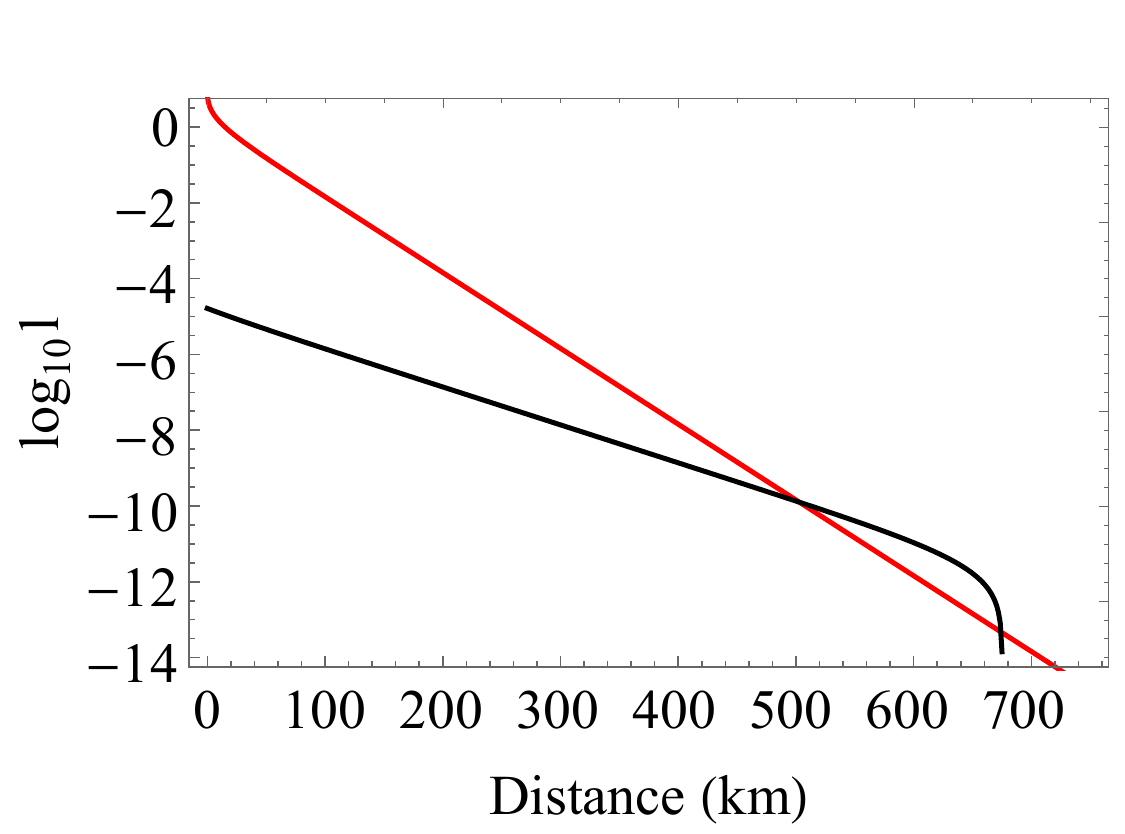}
 \end{center}
 \caption{Log scale (with base 10) of the key rate $l$ as a function of the distance between Alice and Bob. Our key rate is shown with the black solid line, whereas the PLOB bound is represented by the red solid line. Here, we used $\eta_{\rm det}=80\%$, which is a figure
reported by a commercial single-photon detector \cite{scontel}, and $p_{\rm dark}=1.0\times 10^{-11}$, and the efficiency of an error correcting code is assumed to be 1.1.}
 \label{key-rate-8}
\end{figure}

With these parameters, we plot the resulting key rate for a particular choice of an intensity in the Code mode in Fig. \ref{key-rate-8}
(black solid line). In the figure, we fix the mean photon numbers as $\mu_{\rm A}=\mu_{\rm B}=0.0012$, which is almost optimal at 500~km. Importantly, our key rate clearly shows the $\sqrt{\eta}$ scaling property of TF-QKD*, which makes it possible to overcome the SKC limit, represented by the red solid line, after about 
500 km of a standard optical fibre.
For the SKC, we use the bound known as ``PLOB''~\cite{PLO+17}, which is given by $-\log_2(1-e^{-\alpha L_{AB}/10})$ where $L_{AB}$ is the distance between Alice and Bob~\cite{footnote}.

\section{Conclusion}\label{main-conclusion}
In this paper, we prove the information theoretic security of a variant of Twin-Field QKD~\cite{LYD+18} both in the asymptotic and
in the finite key size regime.
Our key idea is to probabilistically switch between the Test mode and Code mode. This way we can exploit the decoy state method
in the Test mode and the phase encoding protocol in the Code mode to generate a key. The use of the decoy state method allows us
to tightly estimate an important parameter that determines the key rate in the Code mode, and we expect a higher key generation rate than a QKD protocol without decoy states.

In fact, our simulation of the key rate in the asymptotic scenario shows that our protocol can indeed outperform the secret key
capacity (SKC) bound using only presently available components.
We plan to complete this analysis by applying our finite size security proof and estimate the number of pulses needed for surpassing the SKC bound.

The attained key rate could be further enhanced in several ways, e.g., by adopting a discrete randomisation~\cite{discrete phase}. of the global phase set by the users, or by removing the requirement of phase randomisation in the Code mode. These and other solutions are currently being investigated and will be the subject of future studies.

\section{Acknowledgment}
K.T. thanks Koji Azuma, Margarida Pereira, and Go Kato for enlightening discussions. K.T. acknowledges support from JST-CREST JPMJCR 1671. H.-K.L. acknowledges fiancial support from the Natural Sciences and Engineering Research Council of Canada (NSERC), the US Office of Naval Research (ONR), Canadian Foundation for Innovation (CFI), Ontario Research Fund (ORF), Huawei Technologies Canada Co., Ltd, and Post-secondary Strategic Infrastructure Fund (SIF).

\newpage

\onecolumngrid
\appendix

\section{Security proof}\label{Appendix-securityproof-asymt}
In this section, we prove the information theoretic security of TF-QKD* protocol. For this, we first introduce a fictitious protocol,
which is mathematically equivalent to TF-QKD* protocol. After we explain an intuition of our security proof,
we move on to the security proof in the asymptotic limit, considering a large number of pulses.
However, in Appendix~\ref{complete proof}, we provide the complete information theoretic security proof using three decoy states in the finite key size.

\subsection{A fictitious protocol for TF-QKD* protocol}\label{Appendix-fictitious}
Like many security proofs of QKD protocols~\cite{review}, it is convenient to convert our TF-QKD${}^{*}$ to an entanglement based
protocol, which we call the fictitious protocol. This protocol provides Eve with exactly the same quantum and classical information
as the actual protocol does, and Eve cannot behave differently between them. Moreover, Alice and Bob's data and data
processing to generate a key is the same as the actual protocol. Therefore, we can employ the fictitious protocol
to prove the security of the actual protocol.

For the construction of the fictitious protocol, we first introduce systems C', C, A, B, E1, and E2, and consider their state
$\ket{\Psi(\theta_{\rm A}, \theta_{\rm B}, \mu)}_{\rm C', C, A, B, E1, E2}$ expressed by

\begin{eqnarray}
&&\ket{\Psi(\theta_{\rm A}, \theta_{\rm B}, \mu_{\rm A}, \mu_{\rm B})}_{\rm C', C, A, B, E1, E2}\nonumber\\
&:=& \ket{0}_{\rm C'} \Big(\sqrt{p_{{\rm Z}_{\rm A}}p_{{\rm Z}_{\rm B}}}\ket{0_{\rm Z}}_{\rm C}\ket{\Psi_{\rm Z_A}(\theta_{\rm A}, \mu_{\rm A})}_{\rm A, E1}\ket{\Psi_{\rm Z_B}(\theta_{\rm B}, \mu_{\rm B})}_{\rm B, E2}+\sqrt{p_{{\rm Y_A}}p_{{\rm Y_B}}}\ket{1_{\rm Z}}_{\rm C}\ket{\Psi_{\rm Y_A}(\theta_{\rm A}, \mu_{\rm A})}_{\rm A, E1}\ket{\Psi_{\rm Y_B}(\theta_{\rm B}, \mu_{\rm B})}_{\rm B, E2}\Big)\nonumber\\
&+& \ket{1}_{\rm C'} \Big(\sqrt{p_{{\rm Z}_{\rm A}}p_{{\rm Y_B}}}\ket{0_{\rm Z}}_{\rm C}\ket{\Psi_{\rm Z_A}(\theta_{\rm A}, \mu_{\rm A})}_{\rm A, E1}\ket{\Psi_{\rm Y_B}(\theta_{\rm B}, \mu_{\rm B})}_{\rm B, E2}+\sqrt{p_{{\rm Y_A}}p_{{\rm Z}_{\rm B}}}\ket{1_{\rm Z}}_{\rm C}\ket{\Psi_{\rm Y_A}(\theta_{\rm A}, \mu_{\rm A})}_{\rm A, E1}\ket{\Psi_{\rm Z_B}(\theta_{\rm B}, \mu_{\rm B})}_{\rm B, E2}\Big)\,,\nonumber\\
\label{starting-total-state}
\end{eqnarray}
where
\begin{eqnarray}
\ket{\Psi_{\rm Z_A}(\theta_{\rm A}, \mu_{\rm A})}_{\rm A, E1}&:=&\frac{1}{\sqrt{2}}\left(\ket{0_{\rm Z}}_{\rm A}(\ket{e^{i \theta_{\rm A}}\sqrt{\mu_{\rm A}}}_{\rm ref}\ket{e^{i \theta_{\rm A}}\sqrt{\mu_{\rm A}}}_{\rm sg})_{\rm E1}+\ket{1_{\rm Z}}_{\rm A}(\ket{e^{i \theta_{\rm A}}\sqrt{\mu_{\rm A}}}_{\rm ref}\ket{e^{i (\theta_{\rm A}+\pi)}\sqrt{\mu_{\rm A}}}_{\rm sg})_{\rm E1}\right),\,\label{Alice's phase state in Z}\\
\ket{\Psi_{\rm Y_A}(\theta_{\rm A}, \mu_{\rm A})}_{\rm A, E1}&:=&\frac{1}{\sqrt{2}}\left(\ket{1_{\rm Y}}_{\rm A}(\ket{e^{i \theta_{\rm A}}\sqrt{\mu_{\rm A}}}_{\rm ref}\ket{e^{i (\theta_{\rm A}+\pi/2)}\sqrt{\mu_{\rm A}}}_{\rm sg})_{\rm E1}+\ket{0_{\rm Y}}_{\rm A}(\ket{\sqrt{e^{i \theta_{\rm A}}\mu_{\rm A}}}_{\rm ref}\ket{e^{i (\theta_{\rm A}+3\pi/2)}\sqrt{\mu_{\rm A}}}_{\rm sg})_{\rm E1}\right)\,,\nonumber\\
\label{Alice's phase state in X}\\
\ket{\Psi_{\rm Z_B}(\theta_{\rm B}, \mu_{\rm B})}_{\rm B, E2}&:=&\frac{1}{\sqrt{2}}\left(\ket{0_{\rm Z}}_{\rm B}(e^{i \theta_{\rm B}}\ket{\sqrt{\mu_{\rm B}}}_{\rm ref}\ket{e^{i \theta_{\rm B}}\sqrt{\mu_{\rm B}}}_{\rm sg})_{\rm E2}+\ket{1_{\rm Z}}_{\rm B}(\ket{e^{i \theta_{\rm B}}\sqrt{\mu_{\rm B}}}_{\rm ref}\ket{e^{i (\theta_{\rm B}+\pi)}\sqrt{\mu_{\rm B}}}_{\rm sg})_{\rm E2}\right),\,\label{Bob's phase state in Z}\\
\ket{\Psi_{\rm Y_B}(\theta_{\rm B}, \mu_{\rm B})}_{\rm B, E2}&:=&\frac{1}{\sqrt{2}}\left(\ket{1_{\rm Y}}_{\rm B}(\ket{e^{i \theta_{\rm B}}\sqrt{\mu_{\rm B}}}_{\rm ref}\ket{e^{i (\theta_{\rm B}+\pi/2)}\sqrt{\mu_{\rm B}}}_{\rm sg})_{\rm E2}+\ket{0_{\rm Y}}_{\rm B}(\ket{e^{i \theta_{\rm B}}\sqrt{\mu_{\rm B}}}_{\rm ref}\ket{e^{i (\theta_{\rm B}+3\pi/2)}\sqrt{\mu_{\rm B}}}_{\rm sg})_{\rm E2}\right)\,.\nonumber\\
\label{Bob's phase state in X}
\end{eqnarray}
In these equations, $\{\ket{0}_{C'}, \ket{1}_{C'}\}$ is the ${\rm Z_{C'}}$ basis for system C', which determines whether
Alice and Bob's bases for state preparations coincide or not,
$\{\ket{0_{\rm Z}}_{C}, \ket{1_{\rm Z}}_{C}\}$ is the ${\rm Z}_{\rm C}$ basis for the quantum coin system C,
and two systems ref and sg correspond to a reference pulse and a signal pulse, respectively.
Alice and Bob send only system sg of systems E1 and E2 to Charlie, who is supposed
to perform a single photon count measurement, while they keep systems C', C, A, B, and ref in their lab.
We have introduced system ref for ease of security proof.
Also, for later convenience, we define the ${\rm X}_{\rm C}$ basis for the quantum coin system C as
$\{\ket{0_{\rm X}}_{\rm C}, \ket{1_{\rm X}}_{\rm C}\}$
where $\ket{0_{\rm X}}_{\rm C}:=\sqrt{p_{\rm Z}^{\rm (AB)}}\ket{0_{\rm Z}}_{C}+\sqrt{p_{\rm Y}^{\rm (AB)}}\ket{1_{\rm Z}}_{C}$
and $\ket{1_{\rm X}}_{\rm C}:=\sqrt{p_{\rm Y}^{\rm (AB)}}\ket{0_{\rm Z}}_{C}-\sqrt{p_{\rm Z}^{\rm (AB)}}\ket{1_{\rm Z}}_{C}$ with
$p_{\rm Z}^{\rm (AB)}:=p_{{\rm Z}_{\rm A}}p_{{\rm Z}_{\rm B}}/(p_{{\rm Z}_{\rm A}}p_{{\rm Z}_{\rm B}}+p_{{\rm Y_A}}p_{{\rm Y_B}})$ and
$p_{\rm Y}^{\rm (AB)}:=p_{{\rm Y_A}}p_{{\rm Y_B}}/(p_{{\rm Z}_{\rm A}}p_{{\rm Z}_{\rm B}}+p_{{\rm Y_A}}p_{{\rm Y_B}})$. Here, notice
that this ${\rm X}_{\rm C}$ is not the standard X basis, and following GLLP~\cite{GLLP}, we have introduced system C
as a quantum coin to take care of Alice's basis choice.
We note that for systems except for systems C, we define the basis states as $\ket{0_{\rm Z}}:=(\ket{0_{\rm X}}+\ket{1_{\rm X}})/\sqrt{2}$
and $\ket{1_{\rm Z}}:=(\ket{0_{\rm X}}-\ket{1_{\rm X}})/\sqrt{2}$ for the Z basis, and $\ket{0_{\rm Y}}:=(\ket{0_{\rm X}}+i
\ket{1_{\rm X}})/\sqrt{2}$ and $\ket{1_{\rm Y}}:=(\ket{0_{\rm X}}-i\ket{1_{\rm X}})/\sqrt{2}$ for the Y basis with $i$ being
the imaginary number, where $\{\ket{0_{\rm X}}, \ket{1_{\rm X}}\}$ is an orthonormal basis. Importantly, we emphasize that the four states in Eqs. (\ref{Alice's phase state in Z})-
(\ref{Bob's phase state in X}) are chosen such that the probability of observing ${\rm X_C}=1$ for ${\rm C'}=0$ is exactly zero
for the emission of the vacuum and a single photon.

With these states, we will represent all of Alice's selections in the actual protocol, including the selection of
the intensity setting and the one of the Test mode or the Code mode, by means of measurements on the following state

\begin{eqnarray}
&&\ket{\Psi}_{\rm {\rm P}_{\rm A}, {\rm P}_{\rm B}, Tes, Int_{A}, Int_{B}, C', C, A, B, E1, E2}:=\sum_{O, \mu_{\rm A}, \mu_{\rm B}}\sum_{n_{\rm A}=0}^{\infty}\sum_{n_{\rm B}=0}^{\infty} \ket{n_{\rm A}}_{\rm {\rm P}_{\rm A}}\ket{n_{\rm B}}_{\rm {\rm P}_{\rm B}}\nonumber\\
&\otimes&\sqrt{p(O,\mu_{\rm A},\mu_{\rm B})}\ket{O}_{\rm Tes}\ket{\mu_{\rm A}}_{\rm Int_{\rm A}}\ket{\mu_{\rm B}}_{\rm Int_{\rm B}}{\hat P}_{n_{\rm A}}^{({\rm E1})}
{\hat P}_{n_{\rm B}}^{({\rm E2})}\ket{\Psi(\theta_{\rm A}, \theta_{\rm B}, \mu_{\rm A}, \mu_{\rm B})}_{\rm C', C, A, B, E1, E2}\,,\nonumber\\
\label{whole state number}
\end{eqnarray}
where $n_{\rm A}$ ($n_{\rm B}$) refers to the photon number contained in systems ref and sg of E1 (E2), $O\in \{{\rm C}, {\rm T}\}$, system Tes is a system to be measured with an orthonormal
basis $\{\ket{\rm T}_{\rm Tes}, \ket{\rm C}_{\rm Tes}\}$ to determines whether it is the Test mode or the Code mode,
and system ${\rm Int}_{\rm A}$ (${\rm Int}_{\rm B}$)
is to be measured with an orthonormal basis
$\{\ket{\mu_{1}}_{\rm Int_{\rm A}}, \ket{\mu_{2}}_{\rm Int_{\rm A}}, \ket{\mu_3}_{\rm Int_{\rm A}}\}$
($\{\ket{\mu_{1}}_{\rm Int_{\rm B}}, \ket{\mu_{2}}_{\rm Int_{\rm B}}, \ket{\mu_3}_{\rm Int_{\rm B}}\}$) to obtain Alice's
(Bob's) intensity setting. Moreover,
$p({\rm C},\mu_{\rm A}, \mu_{\rm B})=0$ for $\mu_{\rm A}\neq \mu_{\rm B}$, $p({\rm T}|\mu_{\rm A}, \mu_{\rm B})=p_{\rm T}$ for $\mu_{\rm A}=\mu_{\rm B}$,
$p({\rm C}|\mu_{\rm A}, \mu_{\rm B})=p_{\rm C}$ for $\mu_{\rm A}=\mu_{\rm B}$, and ${\hat P}_{n}^{({\rm E1})}$
(${\hat P}_{n}^{({\rm E2})}$) is a projection operator to a $n$ photon space of systems ref and sg of E1 (E2).
One can see that the quantum information available to Charlie is the same as the one of the actual protocol if Alice and Bob
send only the signal systems of E1 and E2 to Charlie.
We note that by using Eq. (\ref{identity}), Eq. (\ref{whole state number}) can be rewritten as

\begin{eqnarray}
&&\ket{\Psi}_{\rm {\rm P}_{\rm A}, {\rm P}_{\rm B}, Tes, Int, C', C, A, B, E1, E2}\nonumber\\
&=&\sum_{O, \mu_{\rm A}, \mu_{\rm B}}\frac{1}{2\pi}\int_{-\pi}^{\pi}d\theta_{\rm A}\int_{-\pi}^{\pi}d\theta_{\rm B} \ket{\theta_{\rm A}}_{\rm {\rm P}_{\rm A}} \ket{\theta_{\rm B}}_{\rm {\rm P}_{\rm B}}\sqrt{p(O, \mu_{\rm A}, \mu_{\rm B})}\ket{O}_{\rm Tes}\ket{\mu_{\rm A}}_{\rm Int_{\rm A}}\ket{\mu_{\rm B}}_{\rm Int_{\rm B}}\ket{\Psi(\theta_{\rm A}, \theta_{\rm B}, \mu_{\rm A}, \mu_{\rm B})}_{\rm C', C, A, B, E1, E2}\,.\nonumber\\
\label{whole state2}
\end{eqnarray}
Therefore, if Alice and Bob choose $\{\ket{n_{\rm A}}_{\rm P_{A}}\ket{n_{\rm B}}_{\rm P_{B}}\}$
($\{\ket{\theta_{\rm A}}_{\rm P_{A}}\ket{\theta_{\rm B}}_{\rm P_{B}}\}$) basis to measure systems
${\rm P_A}$ and ${\rm P_B}$, then Alice and Bob
prepare systems sg and ref of E1 and E2 in a photon number state (a non-phase randomized state with the phase).

Most importantly, since Alice and Bob do not disclose the phase information in the Test mode of the actual protocol,
the state of pulses remain exactly the same from Eve or Charliefs viewpoint even if Alice and Bob first prepare
$\ket{\Psi}_{\rm {\rm P}_{\rm A}, {\rm P}_{\rm B}, Tes, Int, C', C, A, B, E1, E2}$, measure the photon number
of systems ${\rm P}_{\rm A}$ and ${\rm P}_{\rm B}$ in the Test mode, and then send only the signal systems of E1 and E2 to Charlie.
This photon number measurement enables Alice and Bob to employ the decoy state method because the state is
a classical mixture of number states. On the other hand, since Alice and Bob announce the phase information in
the Code mode of the actual protocol, this mode is
equivalently described by the preparation of
$\ket{\Psi}_{\rm {\rm P}_{\rm A}, {\rm P}_{\rm B}, Tes, Int, C', C, A, B, E1, E2}$ followed by the phase measurements,
sending only the signal systems of E1 and E2 to Charlie, and the announcement of the phase information. Hence,
Alice and Bob cannot employ the decoy state method in the Code mode because the phase information is leaked to Eve or Charlie,
and the state is no longer regarded as the classical mixture of number states from the viewpoint of Eve or Charlie.

Below, we present how the fictitious protocol runs.
Note that we assume that Alice and Bob are located in the same lab in the fictitious protocol such
that they can exchange some classical information without revealing it to Eve or Charlie,
however we design the protocol in such a way that all the quantum and classical information available to Eve or Charlie
as well as the key generated are the same as those in the actual protocol. Therefore, we can use this protocol
to prove the security.

\bigskip

{\bf \noindent Fictitious protocol}

\begin{enumerate}
\item Alice and Bob repeat Step 2-3, $N$ times. All the public announcements by Alice and Bob are done over an authenticated channel.

\item
Alice and Bob prepare systems ${\rm P}_{\rm A}$, ${\rm P}_{\rm B}$, ${\rm Tes}$, ${\rm Int_{A}}$, ${\rm Int_{B}}$, ${\rm C'}$, ${\rm C}, 
${\rm A}$, ${\rm B}, ${\rm E1}$, and E2 in the state
$\ket{\Psi}_{\rm {\rm P}_{\rm A}, {\rm P}_{\rm B}, Tes, Int_{A}, Int_{B}, C', C, A, B, E1, E2}$ defined in Eq. (\ref{whole state number}).
Then, Alice measures system C' with the $\{\ket{0}_{C'}, \ket{1}_{C'}\}$ basis. Next,
Alice and Bob measure mean photon numbers contained each of systems ${\rm Int_{A}}$ and ${\rm Int_{B}}$, respectively.
Finally, Alice and Bob send only the signal pulses in 
systems  E1 and E2 (see Eqs. (\ref{Alice's phase state in Z})-(\ref{Bob's phase state in X}))
to Charlie over a quantum channel, while they keep all the other systems in the lab.

\item
Charlie performs some measurement on the incoming signals. Ideally, he is supposed to perform a single photon counting
measurement on systems sg of E1 and E2, and if he obtains a double click, then
he randomly decides $1$ or $2$. However, he could do anything he pleases.

If Charlie obtains a detection event, he announces this as well as the type of the outcome $t_{\rm E}\in\{1, 2\}$
over a public channel. If Charlie announces outcomes other than $1$ or $2$, including a non-detection outcome, Alice and Bob discard all the data associated with this event.

\item
For each of the detection events, Alice and Bob announce their intensity selections. Also, depending on whether $\mu_{\rm A}=\mu_{\rm B}$ is satisfied or not, they conduct the following operations:

(i) If $\mu_{\rm A}\neq \mu_{\rm B}$, then
Alice and Bob measure systems ${\rm P_{A}}$ and ${\rm P_{B}}$ with $\{\ket{n_{\rm A}}\}$ and $\{\ket{n_{\rm B}}\}$
bases, respectively, Alice measures system C with the ${\rm Z_C}$ basis, and
Alice and Bob announce their basis choices. That is, Alice announces the ${\rm Z}_{\rm A}$ basis when ${\rm Z_{C}}=0$ and the ${\rm Y_A}$ basis
when ${\rm Z_{C}}=1$, and Bob announces the ${\rm Z}_{\rm B}$ (${\rm Y_B}$) basis when ${\rm Z_{C}}=0$ and C'
outputs 0 (${\rm Z_{C}}=1$
and C' outputs 0) or when ${\rm Z_{C}}=1$ and C' outputs 1 (${\rm Z_{C}}=0$ and C' outputs 1).

(ii) If $\mu_{\rm A}=\mu_{\rm B}$, Alice measures system Tes to determine whether
each of the systems are associated to the Test mode or the Code mode, and announces the outcome.

(ii-i) If the Test mode was selected, then
Alice and Bob measure systems ${\rm P_{A}}$ and ${\rm P_{B}}$ with $\{\ket{n_{\rm A}}\}$ and $\{\ket{n_{\rm B}}\}$
bases, respectively, Alice measures system C with the ${\rm Z_C}$ basis, and
Alice (Bob) announces the basis choice ${\rm Z_A}$ (${\rm Z_B}$) or ${\rm Y_A}$ (${\rm Y_B}$) with the same
manner as in (i).

(ii-ii) If the Code mode was selected, Alice (Bob) measures systems ${\rm P_{A}}$ (${\rm P_{B}}$)
with the phase base $\{\ket{\theta_{\rm A}}\}$ ($\{\ket{\theta_{\rm B}}\}$) and announces the outcome
$\theta_{\rm A}$ ($\theta_{\rm B}$). Next, Alice and Bob measure systems A and B with the ${\rm Y_A}$
and ${\rm Y_B}$ bases, respectively. Then, Alice chooses between ${\rm Z_C}$ and ${\rm X_C}$ with probabilities
$p_{\rm Z_C}$ and $p_{\rm X_C}$, respectively, and announces the selection.

When ${\rm Z_C}$ was selected and $|\theta_{\rm A}-\theta_{\rm B}|\le\Delta/2$ is satisfied,
Alice measures system C with the ${\rm Z_C}$ basis, and Alice (Bob) announces the basis choice
${\rm Z_A}$ (${\rm Z_B}$) or ${\rm Y_A}$ (${\rm Y_B}$) with the same
manner as in (i).
Alice and Bob announce the outcomes of the ${\rm Y_A}$ and ${\rm Y_B}$ bases measurements
only when they announce the ${\rm Y_A}$ and ${\rm Y_B}$ bases.

When ${\rm X_C}$ was selected and $|\theta_{\rm A}-\theta_{\rm B}|\le\Delta/2$ is satisfied,
Alice measures system C with the ${\rm X_C}$ basis, and Alice and Bob announce nothing.

Finally, when $|\theta_{\rm A}-\theta_{\rm B}|\le\Delta/2$ ($|\theta_{\rm A}-\theta_{\rm B}|>\Delta/2$),
Alice and Bob keep (discard) the measurement outcomes. However, they keep the record of the number of the outcomes
occurred even if they discard the data.

\item Alice and Bob announce a small portion of a previously shared secret key (this is done to
simulate the exchange of the encrypted syndrome information in the actual protocol). Then,
Alice selects a hash function randomly according to the result of a parameter estimation based on the data in Step 4, and announces the selected function.
\end{enumerate}

\begin{figure}
\begin{center}
 \includegraphics[scale=0.45]{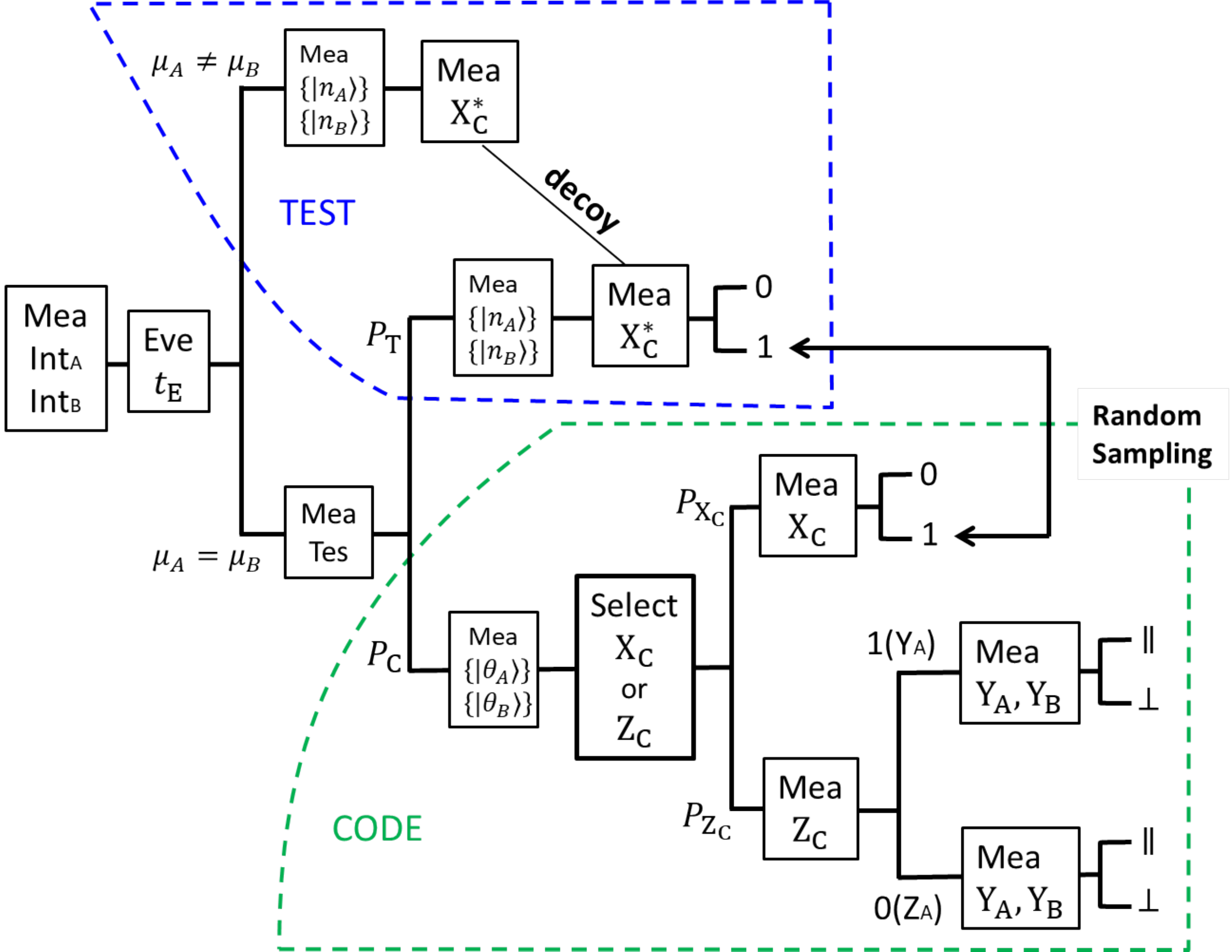}
 \end{center}
 \caption{Logical steps of the fictitious protocol, used to prove the security of the actual protocol. Note that this shows only the case for
$C'=0$, i.e. Alice and Bob's bases used for their state preparations in the fictitious protocol coincide, and all the events with $C'=1$ are not
used in our proof. Here, ``Mea'' represent a measurement,
and $\ket{n}$ ($\ket{\theta}$) refers to a photon number (phase) basis. Each branch represents that we have an outcome in a probabilistic manner.
The superscript $*$ in ${\rm X_C^{*}}$ means that we consider the ${\rm Z_C}$ basis in the fictitious protocol, 
however only for the purpose of estimating parameters needed in our security proof, we are allowed to consider the ${\rm X_C}$ basis instead.
Finally, $||$ ($\perp$) means that the Alice and Bob's measurement outcomes coincide (differ), and
``decoy'' represents that the numbers of instances subjected to the measurement
for each photon number space are estimated by the decoy state method.}
 \label{fig-diagram}
\end{figure}

The logical schematics of the fictitious protocol is shown in Fig.~\ref{fig-diagram}. 
We remark that the ${\rm Z_A}$ and ${\rm Z_B}$ bases have to be used to generate sifted bits
for ${\rm Z_C}=0$ and $C'=0$ in the Code mode. However, we considered to use
${\rm Y_A}$ and ${\rm Y_B}$ bases, complementary observables of the ${\rm Z_A}$ and ${\rm Z_B}$ bases, to measure a phase error.
This is so because in most of the security proofs based on entanglement distillation~\cite{SP},
on the complementary scenario~\cite{koashi09} and on the entropic uncertainty relationship~\cite{Renner}, it is
widely known that we have to estimate the phase error rate $e_{\rm ph}$, which is a fictitious error rate
that we would obtain if we employed the complementary basis for the measurement. This estimated rate
is later to be used in privacy amplification to generate a secure key (more precisely, the fraction of
$h(e_{\rm ph})$, where $h(x)$ is the binary entropy function, has to be sacrificed in privacy amplification in
the limit of large sifted key). As a consequence, the fictitious protocol does not produce a key,
and this is only for estimating phase errors.
However, if needed, a key can be generated if Alice and Bob does not perform the ${\rm Y_A}$ and ${\rm Y_B}$ bases 
measurement on the events with the announcement of the ${\rm Z_A}$ and ${\rm Z_B}$ bases and $|\theta_{\rm A}-\theta_{\rm B}|\le\Delta/2$ in the Code mode, but instead they run an entanglement distillation protocol \cite{SP, EDP} for such events. In this case, we can remove the encryption of the syndrome information for error correction in the actual protocol, and Alice and Bob are allowed to announce the syndrome information without encryption both in
the actual and fictitious protocols. But for simplicity of discussions, we consider not to produce a key.
Final remark on phase errors is;
for $t_{\rm E}=1$ ($t_{\rm E}=2$), a phase error is a coincidence (erroneous) event in the ${\rm Y_A}$ and ${\rm Y_B}$ bases
measurement (see Appendix C for a more detail discussions on the definition of the phase error rate).

A crucial difference of the fictitious protocol from the actual protocol is that Alice and Bob generate a reference pulse, but
they do not send it, whereas Alice and Bob do not even generate such a reference pulse in the actual protocol.
The reason for considering a double pulse is that the bias can be made small, and therefore we can
achieve higher performance.
Importantly, from Eqs. (\ref{starting-total-state})-(\ref{whole state2}), one can see that the statistics of Alice and Bob's observables
that are directly available in the actual
protocol, i.e. Alice and Bob's basis choices,
the bit value choices, the intensity settings, the choice between the Test and Code modes, and the phase information, do not
change with the preparation of the reference pulse. Considering that the reference pulses are not sent to Eve, meaning
that Eve's accessible information remains exactly the same,
we conclude that as long as Alice and Bob's data processing, especially privacy amplification, are the same
between the actual protocol and the fictitious protocol, the security of the actual protocol directly follows from the security
of the fictitious protocol. Hence, we are allowed to focus on the security proof of the fictitious protocol.

Another remark on the fictitious protocol is that measurements on systems
${\rm P_{A}}$, ${\rm P_{B}}$, Tes, ${\rm Int_{A}}$, ${\rm Int_{B}}$, C', C, A, B,
E1, and E2 in the fictitious protocol commute with each other since they are measurements on different systems, and therefore it does not matter which system is measured before or after the other systems. However, we have chosen the order of the measurements
as prescribed. In particular, system C' and systems ${\rm Int_{A}}$ and ${\rm Int_{B}}$ are measured first,
that is, whether Alice and Bob's bases coincide or not and Alice and Bob's intensity settings, are predetermined before Alice and Bob
send the signal systems of E1 and E2.

\subsection{Intuition of our security proof}\label{Sketch-of-the-proof}

Here, we describe an intuition of our proof. As we have discussed, our central problem is to estimate the phase error rate, and
in so doing, we generalize the security proof in~\cite{TLF+12, LP06}. In such a proof, an important quantity is the bias of
the quantum coin (system C), i.e.
the number of ${\rm X_C}=1$ for the events in the Code mode with ${\rm Z_{C'}}=0$ (that is, Alice and Bob's bases selections coincide),
$\mu_{\rm A}=\mu_{\rm B}$,
$|\theta_{\rm A}-\theta_{\rm B}|\le\Delta/2$, and $t_{\rm E}$. Here, note that the bias is defined by the ${\rm X_C}=1$ basis, whereas
our fictitious protocol employs only the ${\rm Z_C}$ basis
for measuring system C in the Test mode. In what follows and throughout the security proof, we consider a Gedanken measurement,
in which we replace all the ${\rm Z_C}$ basis in the {\it Test} mode
with the ${\rm X_C}$ basis, and we will estimate
how many bias we could have obtained if Alice had measured such systems C in the Test mode with the ${\rm X_C}$ basis rather than the
${\rm Z_C}$ basis. Most importantly, as we will see later, this  ${\rm X_C}$ basis measurements correspond to independent trials whose probability can be
readily obtained. Therefore, once we know the number of such instances, which is in fact possible in our protocol thanks to 
the basis announcement made when the ${\rm Z_C}$ is selected, we can readily estimate the number
of ${\rm X_C}=1$ using some probability inequalities, such as Chernoff bound~\cite{Chernoff} or Hoeffding's inequality~\cite{Hoeffding}. 
This is the reason why we are allowed to consider the Gedanken measurement.

Intuitively, the bias represents how differently Eve could behave between the ${\rm Z}$ and ${\rm Y}$ bases states. One
can see this, for instance, by considering
\begin{eqnarray}
&&_{\rm C'}\langle0\ket{\Psi(\theta_{\rm A}, \theta_{\rm B}, \mu_{\rm A}, \mu_{\rm B})}_{\rm C', C, A, B, E1, E2}\nonumber\\
&=&\sqrt{p_{{\rm Z}_{\rm A}}p_{{\rm Z}_{\rm B}}}\ket{0_{\rm Z}}_{\rm C}\ket{\Psi_{\rm Z_A}(\theta_{\rm A}, \mu_{\rm A})}_{\rm A, E1}\ket{\Psi_{\rm Z_B}(\theta_{\rm B}, \mu_{\rm B})}_{\rm B, E2}+\sqrt{p_{{\rm Y_A}}p_{{\rm Y_B}}}\ket{1_{\rm Z}}_{\rm C}\ket{\Psi_{\rm Y_A}(\theta_{\rm A}, \mu_{\rm A})}_{\rm A, E1}\ket{\Psi_{\rm Y_B}(\theta_{\rm B}, \mu_{\rm B})}_{\rm B, E2},\nonumber\\
&=&\ket{0_{\rm X}}_{\rm C}(p_{{\rm Z}_{\rm A}}p_{{\rm Z}_{\rm B}}\ket{\Psi_{\rm Z_A}(\theta_{\rm A}, \mu_{\rm A})}_{\rm A, E1}\ket{\Psi_{\rm Z_B}(\theta_{\rm B}, \mu_{\rm B})}_{\rm B, E2}+p_{{\rm Y}_{\rm A}}p_{{\rm Y}_{\rm B}}\ket{\Psi_{\rm Y_A}(\theta_{\rm A}, \mu_{\rm A})}_{\rm A, E1}\ket{\Psi_{\rm Y_B}(\theta_{\rm B}, \mu_{\rm B})}_{\rm B, E2})\nonumber\\
&+&\sqrt{p_{{\rm Z}_{\rm A}}p_{{\rm Z}_{\rm B}}p_{{\rm Y}_{\rm A}}p_{{\rm Y}_{\rm B}}}\ket{1_{\rm X}}_{\rm C}(\ket{\Psi_{\rm Z_A}(\theta_{\rm A}, \mu_{\rm A})}_{\rm A, E1}\ket{\Psi_{\rm Z_B}(\theta_{\rm B}, \mu_{\rm B})}_{\rm B, E2}-\ket{\Psi_{\rm Y_A}(\theta_{\rm A}, \mu_{\rm A})}_{\rm A, E1}\ket{\Psi_{\rm Y_B}(\theta_{\rm B}, \mu_{\rm B})}_{\rm B, E2})
\label{starting-total-state-later}
\end{eqnarray}
which is obtained from Eq. (\ref{starting-total-state}). From this equation, we observe that if Alice and Bob's state are
the same between the two bases, i.e. they are basis independent, then the probability of obtaining ${\rm X_C}=1$ is exactly zero,
whereas it is not zero for basis dependent states (here recall the definition
$\ket{0_{\rm X}}_{\rm C}:=\sqrt{p_{\rm Z}^{\rm (AB)}}\ket{0_{\rm Z}}_{C}+\sqrt{p_{\rm Y}^{\rm (AB)}}\ket{1_{\rm Z}}_{C}$
and $\ket{1_{\rm X}}_{\rm C}:=\sqrt{p_{\rm Y}^{\rm (AB)}}\ket{0_{\rm Z}}_{C}-\sqrt{p_{\rm Z}^{\rm (AB)}}\ket{1_{\rm Z}}_{C}$
with $p_{\rm Z}^{\rm (AB)}=p_{{\rm Z}_{\rm A}}p_{{\rm Z}_{\rm B}}/(p_{{\rm Z}_{\rm A}}p_{{\rm Z}_{\rm B}}+p_{{\rm Y_A}}p_{{\rm Y_B}})$ and
$p_{\rm Y}^{\rm (AB)}=p_{{\rm Y_A}}p_{{\rm Y_B}}/(p_{{\rm Z}_{\rm A}}p_{{\rm Z}_{\rm B}}+p_{{\rm Y_A}}p_{{\rm Y_B}})$).
Hence, one may presume that the bias in {\it a detection event} has to be small for better key rate because
it becomes difficult for Eve to behave differently between the two bases with a smaller bias (recall that roughly speaking, the data
from one basis monitors a disturbance that Eve caused in the other basis, i.e. the key generation basis).
In the analyses presented in~\cite{TLF+12, LP06}, however, they simply assume the worst case
scenario that by carefully selecting which signals to measure, Eve can
detect signals only for the events with ${\rm X_C}=1$, whereas she does not detect signals
for the events with ${\rm X_C}=0$. This way, an enhancement of the
bias could occur by exploiting channel losses, resulting in a poor key generation rate.

Our key idea to circumvent this worst case scenario is to ask Alice to choose between the Test and Code modes {\it after} Charlie
announces a detection event, as was described in the fictitious protocol. We will show that states conditional on the Test
mode and on the Code mode are exactly the same, following that Charlie or Eve cannot behave differently between the two modes. 
This is natural because Alice sends out the same state between the Code and Test modes, and moreover the choice between the two modes
is made {\it after} Charlie announces his measurement outcome.
These lead us to a random sampling argument that detected events with ${\rm X_C}=1$ is probabilistically assigned, according to the probability
of choosing between the two modes, to the Test or the Code modes {\it after} a detection event (see Sec. \ref{subsection: Fair sampling} for the detail). Here, recall that we consider the Gedanken measurement in which we replace all the ${\rm Z_C}$ basis in the Test mode
with the ${\rm X_C}$ basis.
Now one may deduce that if the bias is small in the
Test mode, so is in the Code mode, and the question is whether the bias in the Test mode is small or not. This is where
the importance of the state selections made in Eqs. (\ref{starting-total-state})-(\ref{Bob's phase state in X}) comes into our analysis.
That is, we have chosen those states such that
the probability of observing ${\rm X_C}=1$ for ${\rm C'}=0$ (where Alice and Bob's bases selections coincide) is exactly zero
for the emission of the vacuum and a single photon, which are dominant contributions to a detection event.
By recalling that Alice and Bob perform the photon number measurement
in the Test mode, we are allowed to consider each photon number space separately,
we may conclude that the bias in the Test mode should be small, resulting in the high key generation rate.

The security proof proceeds as follows; First we rigorously prove the fair sampling argurment. Next, we introduce an inequality
for obtaining phase errors, which is essentially the same as the one presented in~\cite{TLF+12, LP06}. Next,
we present how to estimate the number of ${\rm X_C}=1$ in the Code mode from the one in the Test mode. Then, we employ the decoy state method to estimate the number of
${\rm X_C}=1$ in the Test mode, which is a good estimate of the number of
${\rm X_C}=1$ in the Code mode, and by plugging this quantity into the inequality for obtaining the phase error rate, we
conclude the security proof.

For convenience, below we define
$\xi_{{\rm Code}, t_{\rm E}}^{\mu, \mu, C'=0}$
($\xi_{{\rm Test}, t_{\rm E}}^{\mu_{\rm A}, \mu_{\rm B}, C'=0}$) as a parameter to identify
a set of the event $\{\mu_{\rm A}=\mu, \mu_{\rm B}=\mu, {\rm Z_{C'}}=0, {\rm Code}, t_{\rm E}\}$
($\{\mu_{\rm A}, \mu_{\rm B}, {\rm Z_{C'}}=0, {\rm Test}, t_{\rm E}\}$), and other parameters are defined with a similar
manner. Moreover, we use a notation such as ${\rm X_C}|\xi_{{\rm Test}, t_{\rm E}}^{\mu_{\rm A}, \mu_{\rm B}, C'=0}$
in order to emphasize the Gedanken ${\rm X_C}$ measurement on system C, which plays a central role in our proof.

\subsection{Fair sampling argument}\label{subsection: Fair sampling}
Here we prove the fair sampling argument for the events with ${\rm X_C}=1$ between in the events
$\xi_{{\rm Code}, t_{\rm E}}^{\mu, \mu, C'=0}$ and in those $\xi_{{\rm Test}, t_{\rm E}}^{\mu, \mu, C'=0}$.
For this, we invoke the predetermination property of the fictitious protocol, and we consider
the instances where Alice and Bob obtain ${\rm Z_{C'}}=0$ and
$\mu_{\rm A}=\mu_{\rm A}=\mu$ in Step 1 (here $\mu\in\{\mu_1, \mu_2, \mu_3\}$). The resulting state is given by

\begin{eqnarray}
&& \sum_{n_{\rm A}=0}^{\infty}\sum_{n_{\rm B}=0}^{\infty} \ket{n_{\rm A}}_{\rm {\rm P}_{\rm A}}\ket{n_{\rm B}}_{\rm {\rm P}_{\rm B}}
\otimes\left(\sum_{O\in \{{\rm Test}, {\rm Code}\}}\sqrt{p(O|\mu, \mu)}\ket{O}_{\rm Tes}\right)\nonumber\\
&\otimes&\ket{\mu}_{\rm Int_{\rm A}}
\ket{\mu}_{\rm Int_{\rm B}}{\hat P}_{n_{\rm A}}^{({\rm E1})}
{\hat P}_{n_{\rm B}}^{({\rm E2})}\ket{\Psi(\theta_{\rm A}, \theta_{\rm B}, \mu_{\rm A}, \mu_{\rm B})}_{\rm C', C, A, B, E1, E2}\,.
\end{eqnarray}
Here, importantly, the state of system Tes is decoupled from all the other states, and therefore its measurement outcome, i.e. the choice of the Test mode or the Code mode, is independent of any other outcomes that could be obtained by any measurement
on all the other systems, including Eve's measurement. In other words, the states of pulses conditional on the Test mode and the Code mode are exactly the same,
and Eve cannot behave differently between the two modes.
This means in particular that the events with the ${\rm X_C}=1$ in $\xi_{{\rm Code}, t_{\rm E}}^{\mu, \mu, C'=0}$ and the ones in $\xi_{{\rm Test}, t_{\rm E}}^{\mu, \mu, C'=0}$ (the Gedanken measurement) are sampled with probabilities
$p({\rm Code}|\mu, \mu)=p_{\rm C}$ and $p({\rm Test}|\mu, \mu)=p_{\rm T}$, respectively, which concludes our fair sampling argument.

\subsection{Security of the Code mode and formula for the key generation length in the asymptotic limit}
In this subsection, we establish the inequality for obtaining the number of phase errors.
The starting point is to recall the commutation property, i.e. measurements on systems ${\rm P_{A}}$, ${\rm P_{B}}$, Tes,
${\rm Int_{A}}$, ${\rm Int_{B}}$, C', C, A, B, E1, and E2 in the fictitious protocol commute with each other. With this property,
we are allowed to imagine that Alice and Bob finish the measurements on systems C, A, and
B to obtain the outcomes of
${\rm Z_{C'}}=0$, $\mu_{\rm A}=\mu_{\rm B}=\mu$, and $|\theta_{\rm A}-\theta_{\rm B}|\le\Delta/2$
before they send the signal systems of E1 and E2.
This instance can equivalently be represented by the following state

\begin{eqnarray}
&&\ket{\Psi(\theta_{\rm A}, \theta_{\rm B}, \mu, \mu)}_{\rm C, A, B, E1, E2}\nonumber\\
&:=&\sqrt{p_{\rm Z}^{\rm (AB)}}\ket{0_{\rm Z}}_{\rm C}\ket{\Psi_{\rm Z_A}(\theta_{\rm A}, \mu)}_{\rm A, E1}\ket{\Psi_{\rm Z_B}(\theta_{\rm B}, \mu)}_{\rm B, E2}+\sqrt{p_{\rm Y}^{\rm (AB)}}\ket{1_{\rm Z}}_{\rm C}\ket{\Psi_{\rm Y_A}(\theta_{\rm A}, \mu)}_{\rm A, E1}\ket{\Psi_{\rm Y_B}(\theta_{\rm B}, \mu)}_{\rm B, E2}\,,\nonumber\\
\end{eqnarray}
with $\mu_{\rm A}=\mu_{\rm B}=\mu$ and $|\theta_{\rm A}-\theta_{\rm B}|\le\Delta/2$, and then Eve or Charlie applies some operations
on systems E1 and E2. In particular, we imagine that Charlie announces $t_{\rm E}$ as her outcome.
We remark that any correlations that Eve or Charlie could cause between this state and states associated to
all the other measurement outcomes can be properly taken into account through
the use of the Azuma's inequality. This is so because this inequality is valid even under any correlations \cite{Three, SCIC}.
Therefore, we are allowed to concentrate only on the preparation of this state and consider Charlie's action on this state.

In order to consider the phase error rate, we consider the Bloch sphere bound, and by applying the Azuma's inequality
we have in the asymptotic limit that (see Eq. (\ref{azuma-number-appendix}) in Appendix \ref{Appendix: phase-number} where we also present the inequality in the finite key size regime)
\begin{eqnarray}
&&N_{{\rm Z_C}|\xi_{{\rm Code}, t_{\rm E}}^{\mu, \mu, C'=0}, \le\Delta/2}-2\frac{p_{\rm Z_C}}{p_{\rm X_C}}N_{{\rm X_C}=1, {\rm X_C}|\xi_{{\rm Code}, t_{\rm E}}^{\mu, \mu, C'=0}}\nonumber\\
&\le&2(p_{\rm Z}^{\rm (AB)}-p_{\rm Y}^{\rm (AB)})\left(p_{{\rm Z_C}}N_{\xi_{{\rm Code}, t_{\rm E}}^{\mu, \mu, C'=0}, \le\Delta/2}-N_{{\rm Y_A, {\rm Z_{C}}}|\xi_{{\rm Code}, t_{\rm E}}^{\mu, \mu, C'=0}, \le\Delta/2}\right)\nonumber\\
&+& 4\sqrt{p_{\rm Z}^{\rm (AB)}p_{\rm Y}^{\rm (AB)}}\sqrt{N_{{Y_{\perp}, \rm Y_A, {\rm Z_{C}}}|\xi_{{\rm Code}, t_{\rm E}}^{\mu, \mu, C'=0}, \le\Delta/2}N_{{Y_{\perp}, \rm Z_A, {\rm Z_{C}}}|\xi_{{\rm Code}, t_{\rm E}}^{\mu, \mu, C'=0}, \le\Delta/2}}\nonumber\\
&+& 4\sqrt{p_{\rm Z}^{\rm (AB)}p_{\rm Y}^{\rm (AB)}}\sqrt{N_{{Y_{||}, \rm Y_A, {\rm Z_{C}}}|\xi_{{\rm Code}, t_{\rm E}}^{\mu, \mu, C'=0}, \le\Delta/2}N_{{Y_{||}, \rm Z_A, {\rm Z_{C}}}|\xi_{{\rm Code}, t_{\rm E}}^{\mu, \mu, C'=0}, \le\Delta/2}}\,,
\label{azuma-number-asympt}
\end{eqnarray}
Here, we consider that this bias is enhanced due to the post-selection depending on whether $|\theta_{\rm A}-\theta_{\rm B}|\le\Delta/2$ or not, which is reflected by
$N_{{\rm X_C}=1, {\rm X_C}|\xi_{{\rm Code},
t_{\rm E}}^{\mu, \mu, C'=0}}\ge N_{{\rm X_C}=1, {\rm X_C}|\xi_{{\rm Code},
t_{\rm E}}^{\mu, \mu, C'=0}, \le\Delta/2}$ where $N_{{\rm X_C}=1, {\rm X_C}|\xi_{{\rm Code},
t_{\rm E}}^{\mu, \mu, C'=0}, \le\Delta/2}$ is the number of the events ${\rm X_C}=1$ and ${\rm X_C}$
among the events specified by $\xi_{{\rm Code}, t_{\rm E}}^{\mu, \mu, C'=0}$ and $\le\Delta/2$. Moreover, $N_{{Y_{\perp}, \rm Y_A, {\rm Z_{C}}}|\xi_{{\rm Code}, t_{\rm E}}^{\mu, \mu, C'=0}, \le\Delta/2}$ is the number
of the events where Alice selects the ${\rm Y_A}$ basis for measuring systems A, ${\rm Z_C}$ is selected, and a Y basis error
occurs among the events $\xi_{{\rm Code}, t_{\rm E}}^{\mu, \mu, C'=0, \le\Delta/2}$. Other numbers in the inequality are defined in a similar manner. Note that the subscript ${||}$ means that the outcome of the ${\rm Y}$ basis measurements coincide, and the inequality in~Eq. (\ref{azuma-number-asympt}) can
be simplified to the inequality as Eq. (\ref{main-simplified-ineq-phase}) in the main text when
$p_{\rm Z_A}=p_{\rm Y_A}=p_{\rm Z_B}=p_{\rm Y_B}$.


In this inequality, $N_{{Y_{\perp}, \rm Y_A, {\rm Z_{C}}}|\xi_{{\rm Code}, t_{\rm E}}^{\mu, \mu, C'=0}, \le\Delta/2}$ and $N_{{Y_{||}, \rm Y_A, {\rm Z_{C}}}|\xi_{{\rm Code}, t_{\rm E}}^{\mu, \mu, C'=0}, \le\Delta/2}$
are quantities we can obtain in the actual protocol, and the important number, i.e. the one of phase errors
for $t_{\rm E}=1$ is given by
$N_{{Y_{\perp}, \rm Z_A, {\rm Z_{C}}}|\xi_{{\rm Code}, t_{\rm E}=1}^{\mu, \mu, C'=0}, \le\Delta/2}:=N_{{Y_{\perp}, \rm Y_A, {\rm Z_{C}}}|\xi_{{\rm Code}, t_{\rm E}=1}^{\mu, \mu, C'=0}, \le\Delta/2}$,
and the one for $t_{\rm E}=2$ is given by
$N_{{Y_{||}, \rm Y_A, {\rm Z_{C}}}|\xi_{{\rm Code}, t_{\rm E}=2}^{\mu, \mu, C'=0}, \le\Delta/2}:=N_{{Y_{||}, \rm Y_A, {\rm Z_{C}}}|\xi_{{\rm Code}, t_{\rm E}=2}^{\mu, \mu, C'=0}, \le\Delta/2}$
(recall the discussion on the definition of a phase error in Appendix C).

For obtaining the upper bound of the number of phase errors, we need to know the number
$N_{{\rm X_C}=1, {\rm X_C}|\xi_{{\rm Code}, t_{\rm E}}^{\mu, \mu, C'=0}}$, however Alice and Bob do not have
a direct access to this number in the actual protocol. Therefore, we have to estimate this number, and we denote
its upper bound by ${\overline N}_{{\rm X_C}=1, {\rm X_C}|\xi_{{\rm Code}, t_{\rm E}}^{\mu, \mu, C'=0}}$.
As we have explained, this number will be estimated via the random sampling theory from the number ${\rm X_C}=1$
that Alice could have obtained if she had chosen the ${\rm X_C}$ basis in the Test mode.
This number and its upper bound are denoted by $N_{{\rm X_C}=1, {\rm X_C}|\xi_{{\rm Test}, t_{\rm E}}^{\mu, \mu, C'=0}}$, and
$\overline{N}_{{\rm X_C}=1, {\rm X_C}|\xi_{{\rm Test}, t_{\rm E}}^{\mu, \mu, C'=0}}$, respectively.
We will present this estimation in the following subsections.

Given the upper bound of the number of phase errors, the key length $l$ is expressed as~\cite{key length, SCIC}

\begin{eqnarray}
l_{\mu, t_{\rm E}=1}&=&N_{{\rm Z_A, {\rm Z_{C}}}|\xi_{{\rm Code}, t_{\rm E}=1}^{\mu, \mu, C'=0}, \le\Delta/2}\left[1-
h\left(\frac{\overline{N}_{{Y_{\perp}, \rm Z_A, {\rm Z_{C}}}|\xi_{{\rm Code}, t_{\rm E}=1}^{\mu, \mu, C'=0}, \le\Delta/2}}{N_{{\rm Z_A, {\rm Z_{C}}}|\xi_{{\rm Code}, t_{\rm E}=1}^{\mu, \mu, C'=0}, \le\Delta/2}}\right)\right]-\lambda_{{\rm EC}, \mu}\,,\\
l_{\mu, t_{\rm E}=2}&=&N_{{\rm Z_A, {\rm Z_{C}}}|\xi_{{\rm Code}, t_{\rm E}=2}^{\mu, \mu, C'=0}, \le\Delta/2}\left[1-
h\left(\frac{\overline{N}_{{Y_{||}, \rm Z_A, {\rm Z_{C}}}|\xi_{{\rm Code}, t_{\rm E}=2}^{\mu, \mu, C'=0}, \le\Delta/2}}{N_{{\rm Z_A, {\rm Z_{C}}}|\xi_{{\rm Code}, t_{\rm E}=2}^{\mu, \mu, C'=0}, \le\Delta/2}}\right)\right]-\lambda_{{\rm EC}, \mu} \,,
\end{eqnarray}
where, $h(x)$ is the binary entropy function, and $\lambda_{{\rm EC}, \mu}$ is the amount of information exchanged for error correction.
In the next section, we explain the estimation of
${\overline N}_{{\rm X_C}=1, {\rm X_C}|\xi_{{\rm Test}, t_{\rm E}}^{\mu, \mu, C'=0}}$ in the following sections.


\subsection{Estimation of $N_{{\rm X_C}=1, {\rm X_C}|\xi_{{\rm Code}, t_{\rm E}}^{\mu, \mu, C'=0}}$ from
$N_{{\rm X_C}=1, {\rm X_C}|\xi_{{\rm Test}, t_{\rm E}}^{\mu, \mu, C'=0}}$}\label{main-estimation-test-code in the asymptotic limit}

In this section, we explain the estimation of $N_{{\rm X_C}=1, {\rm X_C}|\xi_{{\rm Code}, t_{\rm E}}^{\mu, \mu, C'=0}}$.
First, recall the discussion in Sec. \ref{subsection: Fair sampling} that the choice between the Code and the Test modes
within the events ${\rm Z_{C'}}=0$ and $\mu_{\rm A}=\mu_{\rm A}=\mu$ is independent of any other events, and we employ this argument in estimating
$\overline{N}_{{\rm X_C}=1, {\rm X_C}|\xi_{{\rm Code}, t_{\rm E}}^{\mu, \mu, C'=0}}$
from $N_{{\rm X_C}=1, {\rm X_C}|\xi_{{\rm Test}, t_{\rm E}}^{\mu, \mu, C'=0}}$. For this, observe that $N_{{\rm X_C}=1, {\rm X_C}|\xi_{{\rm Code}, t_{\rm E}}^{\mu, \mu, C'=0}}$
and $N_{{\rm X_C}=1, {\rm X_C}|\xi_{{\rm Test}, t_{\rm E}}^{\mu, \mu, C'=0}}$ remain unchanged even if we perform the ${\rm X_{C}}$ basis measurement
on systems C in the Code mode with the selection of ${\rm Z_C}$ basis. This is so because measurements
on different systems commute. Therefore, only for the purpose for
estimating $N_{{\rm X_C}=1, {\rm X_C}|\xi_{{\rm Code}, t_{\rm E}}^{\mu, \mu, C'=0}}$
from $N_{{\rm X_C}=1, {\rm X_C}|\xi_{{\rm Test}, t_{\rm E}}^{\mu, \mu, C'=0}}$, we are allowed to suppose that
Alice measures systems C with the ${\rm X_{C}}$ basis, and then each of the instances with ${\rm X_{C}}=1$ is assigned either
to the Test mode or to the selection of ${\rm X_C}$ basis in the Code mode with probabilities $s_{{\rm X}_{\rm C}}$ and $1-s_{{\rm X}_{\rm C}}$, respectively.
Here, $s_{{\rm X}_{\rm C}}:=p_{\rm T}/(p_{\rm T}+p_{\rm C}p_{{\rm X}_{\rm C}})$.
That is, we have $N_{{\rm X_C}=1, {\rm X_C}|\xi_{{\rm Code}, t_{\rm E}}^{\mu, \mu, C'=0}}+N_{{\rm X_C}=1, {\rm X_C}|\xi_{{\rm Test}, t_{\rm E}}^{\mu, \mu, C'=0}}$ of 1's and
these 1's are assigned either to the Test or Code modes with the Bernoulli trials, and we have that
\begin{eqnarray}
N_{{\rm X_C}=1, {\rm X_C}|\xi_{{\rm Code}, t_{\rm E}}^{\mu, \mu, C'=0}}= \frac{1-s_{{\rm X}_{\rm C}}}{s_{{\rm X}_{\rm C}}}N_{{\rm X_C}=1, {\rm X_C}|\xi_{{\rm Test}, t_{\rm E}}^{\mu, \mu, C'=0}}\nonumber\\
\label{estimate-code-test-main}
\end{eqnarray}
holds.
Next problem is to estimate $N_{{\rm X_C}=1, {\rm X_C}|\xi_{{\rm Test}, t_{\rm E}}^{\mu,
\mu, C'=0}}$ by using the decoy state method, which we present in the next section. 

\subsection{Estimation of $N_{{\rm X_C}=1, {\rm X_C}|\xi_{{\rm Test}, t_{\rm E}}^{\mu,
\mu, C'=0}}$ using the decoy state method in the asymptotic limit}\label{subsection-decoy}

In this section, we present how to estimate $N_{{\rm X_C}=1, {\rm X_C}|\xi_{{\rm Test}, t_{\rm E}}^{\mu,
\mu, C'=0}}$.
For this, recall that in the Test mode
systems ${\rm P}_{\rm A}$ and ${\rm P}_{\rm B}$ are measured with the photon number basis, and therefore states of composite
systems of the signal and reference pulses in E1 and E2 are classical mixtures of photon number states.
Therefore, we can decompose $N_{{\rm X_C}=1, {\rm X_C}|\xi_{{\rm Test}, t_{\rm E}}^{\mu, \mu, C'=0}}$
into
\begin{eqnarray}
N_{{\rm X_C}=1, {\rm X_C}|\xi_{{\rm Test}, t_{\rm E}}^{\mu, \mu, C'=0}}=
\sum_{n_{\rm A}, n_{\rm B}}N_{{\rm X_C}=1, {\rm X_C}|\xi_{{\rm Test}, t_{\rm E}}^{\mu, \mu, C'=0}, n_{\rm A}, n_{\rm B}}\,.\nonumber\\
\label{d-number-main}
\end{eqnarray}
Here, $N_{{\rm X_C}=1, {\rm X_C}|\xi_{{\rm Test}, t_{\rm E}}^{\mu, \mu, C'=0}, n_{\rm A}, n_{\rm B}}$ is the number of the events with
${\rm X_C}=1$ and the selection of ${\rm X_C}$ among the
events where the event specified by $\xi_{{\rm Test}, t_{\rm E}}^{\mu, \mu, C'=0}$ occurred, and Alice and Bob respectively emitted $n_{\rm A}$ and $n_{\rm B}$ photons. We define $N_{{\rm X_C}=1, {\rm X_C}|\xi_{{\rm Test}, t_{\rm E}}^{\mu_{\rm A}, \mu_{\rm B}, C'=0}, n_{\rm A}, n_{\rm B}}$ in the same manner.
Here, recall that the subscript ${\rm X_C}|$ is to emphasize the Gedanken measurement,
in which we replace all the ${\rm Z_C}$ basis in the Test mode 
with the ${\rm X_C}$ basis. Next, in order to compute $N_{{\rm X_C}=1, {\rm X_C}|\xi_{{\rm Test}, t_{\rm E}}^{\mu, \mu, C'=0}}$, we further decompose Eq. (\ref{d-number-main}) into 

\begin{eqnarray}
N_{{\rm X_C}=1, {\rm X_C}|\xi_{{\rm Test}, t_{\rm E}}^{\mu, \mu, C'=0}}&=&
\sum_{n_{\rm A}, n_{\rm A}|(n_{\rm A}, n_{\rm B})\notin \{(0,0), (1,0), (0,1), (1,1)\}}N_{{\rm X_C}=1, {\rm X_C}|\xi_{{\rm Test}, t_{\rm E}}^{\mu, \mu, C'=0}, n_{\rm A}, n_{\rm B}}\nonumber\\
&\le& \sum_{n_{\rm A}, n_{\rm A}|(n_{\rm A}, n_{\rm B})\notin \{(0,0), (1,0), (0,1), (1,1)\}}N_{{\rm X_C}|\xi_{{\rm Test}, t_{\rm E}}^{\mu, \mu, C'=0}, n_{\rm A}, n_{\rm B}}\,.\label{first eq-main}
\end{eqnarray}
where $N_{{\rm X_C}|\xi_{{\rm Test}, t_{\rm E}}^{\mu, \mu, C'=0}, n_{\rm A}, n_{\rm B}}$ is the same as the number of
events with $n_{\rm A}$ and $n_{\rm B}$ photons emitted and $\xi_{{\rm Test}, t_{\rm E}}^{\mu, \mu, C'=0}$ in the fictitious protocol.
Here, the inequality is due to the fact that the event specified by ${\rm X_C}=1$ is
a subset of the one specified by ${\rm X_C}=0\cup {\rm X_C}=1$, which is denoted by ${\rm X_C}$.
In Eq. (\ref{first eq-main}), we have used the fact that
we have chosen the states in Eqs. (\ref{Alice's phase state in Z})-(\ref{Bob's phase state in X}) such that
the probability of observing ${\rm X_C}=1$ for ${\rm C'}=0$ is exactly zero for $(n_{\rm A}, n_{\rm B})\in \{(0,0), (1,0), (0,1), (1,1)\}$
(see Appendix \ref{zero-pro} for more detail). Eq. (\ref{first eq-main}) means that once we can estimate the number $N_{{\rm X_C}|\xi_{{\rm Test}, t_{\rm E}}^{\mu, \mu, C'=0}, n_{\rm A}, n_{\rm B}}$, then we can estimate the quantity of our interest that we do not measure in the fictitious protocol.
Importantly, in estimating $N_{{\rm X_C}|\xi_{{\rm Test}, t_{\rm E}}^{\mu, \mu, C'=0}, n_{\rm A}, n_{\rm B}}$, it does not matter which basis we use for the measurements. This confirms the justification of the use of the Gedanken measurement.

From Eqs. (\ref{first eq-main}), one can see that our problem is to estimate $N_{{\rm X_C}|\xi_{{\rm Test}, t_{\rm E}}^{\mu, \mu, C'=0}, n_{\rm A}, n_{\rm B}}$.
For this, recall the standard decoy state argument that when Alice and Bob respectively emit $n_{\rm A}$ and $n_{\rm B}$ photons
to Charlie, those photons do not contain any information about the intensity setting. This is so because we assume that there is no
state preparation flaw and side channel. Therefore, one can imagine that
Alice and Bob perform the photon number measurements first, and then they probabilistically assign their intensity settings
{\it after} Charlie announces his detection result $t_{\rm E}$. With this observation, we have
\begin{eqnarray}
\sum_{n_{\rm A}, n_{\rm B}} N_{n_{\rm A}, n_{\rm B}|\xi_{{\rm Test}, t_{\rm E}}^{C'=0}}q_{\mu_{\rm A}, \mu_{\rm B}|n_{\rm A}, n_{\rm B}}
=N_{{\rm X_C}|\xi_{{\rm Test}, t_{\rm E}}^{\mu_{\rm A}, \mu_{\rm B}, C'=0}}\,,\label{hof3}
\end{eqnarray}
where $q_{\mu_{\rm A}, \mu_{\rm B}|n_{\rm A}, n_{\rm B}}$,
is a probability that Alice and Bob respectively select an intensity setting $\mu_{\rm A}$ and $\mu_{\rm B}$,
given that Alice and Bob respectively emit $n_{\rm A}$ and $n_{\rm B}$ photons (the explicit form of
$q_{\mu_{\rm A}, \mu_{\rm B}|n_{\rm A}, n_{\rm B}}$ is given in Appendix \ref{explicit-form}).
Thanks to the assumption of the infinite decoy states, which we have assumed for simplicity of discussions, we can obtain
$N_{n_{\rm A}, n_{\rm B}|\xi_{{\rm Test}, t_{\rm E}}^{C'=0}}$ using the experimentally available data.
After obtaining this, we consider to probabilistically assign intensity settings to those photon number instances
to have
\begin{eqnarray}
N_{{\rm X_C}|\xi_{{\rm Test}, t_{\rm E}}^{\mu, \mu, C'=0}, n_{\rm A}, n_{\rm B}}=N_{n_{\rm A}, n_{\rm B}|\xi_{{\rm Test}, t_{\rm E}}^{C'=0}}q_{\mu, \mu|n_{\rm A}, n_{\rm B}}\,.
\end{eqnarray}
By substituting this into Eq. (\ref{first eq-main})
we can express $N_{{\rm X_C}=1, {\rm X_C}|\xi_{{\rm Code}, t_{\rm E}}^{\mu, \mu, C'=0}}$ from
$N_{{\rm X_C}=1, {\rm X_C}|\xi_{{\rm Test}, t_{\rm E}}^{\mu, \mu, C'=0}}$
through Eq. (\ref{estimate-code-test-main}). This concludes the security proof in the asymptotic limit.

\section{Relationship between the phase states and the number states}\label{phase-number-relationship}

In this Appendix, we prove Eq. (\ref{identity}). For this, first, we show the identity $\langle m\ket{n}=\delta_{m,n}$ as

\begin{eqnarray}
\langle m\ket{n}=\frac{1}{2\pi}\int_{0}^{2\pi}\int_{0}^{2\pi}d\theta'd\theta e^{i(-m\theta'+n\theta)}\delta (\theta'-\theta)=\frac{1}{2\pi}\int_{0}^{2\pi}d\theta' e^{i(-m+n)\theta'}=\delta_{m,n}\,.
\end{eqnarray}

Next, we prove Eq. (\ref{identity}). Let us define
\begin{eqnarray}
\ket{\Psi}_{\rm P, B}:=\frac{1}{\sqrt{2\pi}}\int_{0}^{2\pi} d\theta \ket{\theta}_{\rm P}\ket{e^{i\theta}\sqrt{\mu}}_{\rm B}\,,
\end{eqnarray}
and we calculate ${}_{\rm P}\langle n\ket{\Psi}_{\rm P, B}$ using Eq.~(\ref{number state}) to obtain
\begin{eqnarray}
{}_{\rm P}\langle n\ket{\Psi}_{\rm P, B}&=&\frac{1}{2\pi}\int_{0}^{2\pi}\int_{0}^{2\pi}d\theta'd\theta e^{-i n\theta}{}_{\rm P}\langle\theta'\ket{\theta}_{\rm P}\ket{e^{i\theta}\sqrt{\mu}}_{\rm B}\nonumber\\
&=&\frac{1}{2\pi}\int_{0}^{2\pi}d\theta' e^{-i n\theta'}\ket{e^{i\theta'}\sqrt{\mu}}_{\rm B}\nonumber\\
&=&e^{-\mu/2}\sum_{m=0}^{\infty}\frac{\sqrt{\mu}^{m}}{\sqrt{m!}} \left(\frac{1}{2\pi}\int_{0}^{2\pi}d\theta'e^{-i(n-m)\theta'}\right)\ket{m}_{\rm B}\nonumber\\
&=& e^{-\mu/2}\frac{\sqrt{\mu}^{n}}{\sqrt{n!}} \ket{n}_{\rm B}\,.
\end{eqnarray}
Therefore, by noting that $\sum_{n=0}^{\infty}\ket{n}_{\rm P}\bra{n}={\hat \openone}_{\rm P}$, we have the relationship as
\begin{eqnarray}
\ket{\Psi}_{\rm P, B}={\hat \openone}_{\rm P} \ket{\Psi}_{\rm P, B}=e^{-\mu/2}\sum_{n=0}^{\infty} \frac{\sqrt{\mu}^{n}}{\sqrt{n!}}\ket{n}_{\rm P} \ket{n}_{\rm B}\,,
\end{eqnarray}
which concludes the proof.

\section{Definition of a phase error}\label{Appendix: def of Ph}

In this section, we consider a situation in which Charlie behaves honestly, that is, we see how the state evolves
when there is no channel losses and noises and Charlie performs a single photon counting measurement.
For this, we present the relationship between an input state and an output state of Charlie's beam splitter as
\begin{eqnarray}
\ket{\alpha}_{\rm E1}\ket{\beta}_{\rm E2}\rightarrow\ket{(\alpha+\beta)/\sqrt{2}}_{\rm E1'}\ket{(\alpha-\beta)/\sqrt{2}}_{\rm E2'}\,.
\end{eqnarray}
Here, E1' and E2' denote the output modes of the beam splitter, and $\alpha$ and $\beta$ are complex numbers
for representing coherent states. We define that Charlie announces $t_{\rm E}=1$ ($t_{\rm E}=2$) when he observes a
detection event only in E1' (E2'), and he announces the non-detection event for all the other cases.
With this relationship, one can see up to the normalization factor that
\begin{eqnarray*}
&&\ket{\Psi_{\rm Z_A}(\theta, \mu)}_{\rm A, E1}\ket{\Psi_{\rm Z_B}(\theta, \mu)}_{\rm B, E2}\\
&\rightarrow& \ket{0_{\rm Z}}_{\rm A}\ket{0_{\rm Z}}_{\rm B}\ket{\sqrt{2\mu}}_{\rm E1'}\ket{0}_{\rm E2'}
+\ket{1_{\rm Z}}_{\rm A}\ket{1_{\rm Z}}_{\rm B}\ket{-\sqrt{2\mu}}_{\rm E1'}\ket{0}_{\rm E2'}\\
&+&\ket{0_{\rm Z}}_{\rm A}\ket{1_{\rm Z}}_{\rm B}\ket{0}_{\rm E1'}\ket{\sqrt{2\mu}}_{\rm E2'}
+\ket{1_{\rm Z}}_{\rm A}\ket{0_{\rm Z}}_{\rm B}\ket{0}_{\rm E1'}\ket{-\sqrt{2\mu}}_{\rm E2'}\,,
\end{eqnarray*}
and 
\begin{eqnarray*}
&&\ket{\Psi_{\rm Y_A}(\theta, \mu)}_{\rm A, E1}\ket{\Psi_{\rm Y_B}(\theta, \mu)}_{\rm B, E2}\\
&\rightarrow& \ket{0_{\rm Y}}_{\rm A}\ket{0_{\rm Y}}_{\rm B}\ket{-i\sqrt{2\mu}}_{\rm E1'}\ket{0}_{\rm E2'}+\ket{1_{\rm Y}}_{\rm A}\ket{1_{\rm Y}}_{\rm B}\ket{i\sqrt{2\mu}}_{\rm E1'}\ket{0}_{\rm E2'}\\
&+&\ket{0_{\rm Y}}_{\rm A}\ket{1_{\rm Y}}_{\rm B}\ket{0}_{\rm E1'}\ket{-i\sqrt{2\mu}}_{\rm E2'}+\ket{1_{\rm Y}}_{\rm A}\ket{0_{\rm Y}}_{\rm B}\ket{0}_{\rm E1'}\ket{i\sqrt{2\mu}}_{\rm E2'}\,,
\end{eqnarray*}
where we consider the case with $\theta_{\rm A}=\theta_{\rm B}=\theta$ and $\mu_{\rm A}=\mu_{\rm B}=\mu$ for simplicity.
These equations suggest that when Charlie observes a single-photon in the event $t_{\rm E}=1$, we obtain the state
$\ket{0_{\rm Z}}_{\rm A}\ket{0_{\rm Z}}_{\rm B}-\ket{1_{\rm Z}}_{\rm A}\ket{1_{\rm Z}}_{\rm B}$ from the Z basis
and $\ket{0_{\rm Y}}_{\rm A}\ket{0_{\rm Y}}_{\rm B}-\ket{1_{\rm Y}}_{\rm A}\ket{1_{\rm Y}}_{\rm B}$ from the Y basis.
On the other hand, when Charlie observes a single-photon in the event $t_{\rm E}=2$, we obtain the state
$\ket{0_{\rm Z}}_{\rm A}\ket{1_{\rm Z}}_{\rm B}-\ket{1_{\rm Z}}_{\rm A}\ket{0_{\rm Z}}_{\rm B}$ from the Z basis
and $\ket{0_{\rm Y}}_{\rm A}\ket{1_{\rm Y}}_{\rm B}-\ket{1_{\rm Y}}_{\rm A}\ket{0_{\rm Y}}_{\rm B}$ from the Y basis.
From this, one can see that Alice and Bob obtain the same bit value if Alice flips her bit value only when Charlie
announces $t_{\rm E}=2$.

Next, we consider how we should define the phase error. For this, we first consider $t_{\rm E}=1$. Note that
$\ket{0_{\rm Z}}_{\rm A}\ket{0_{\rm Z}}_{\rm B}-\ket{1_{\rm Z}}_{\rm A}\ket{1_{\rm Z}}_{\rm B}$ can
be rewritten as $\ket{0_{\rm Y}}_{\rm A}\ket{0_{\rm Y}}_{\rm B}-\ket{1_{\rm Y}}_{\rm A}\ket{1_{\rm Y}}_{\rm B}$, where we have used
$\ket{0_{\rm Z}}=e^{-i\pi/4}(\ket{0_{\rm Y}}+i\ket{1_{\rm Y}})/\sqrt{2}$ and
$\ket{1_{\rm Z}}=e^{i\pi/4}(\ket{0_{\rm Y}}-i\ket{1_{\rm Y}})/\sqrt{2}$. This may lead us to a conclude
that for $t_{\rm E}=1$, we adopt the definition of
the phase error such that it is an erroneous event in Alice and Bob's fictitious Y basis measurements given the Z basis state
preparation.

Similarly, as for $t_{\rm E}=2$, by noting that $\ket{0_{\rm Z}}_{\rm A}\ket{1_{\rm Z}}_{\rm B}-\ket{1_{\rm Z}}_{\rm A}\ket{0_{\rm Z}}_{\rm B}$ can
be rewritten as $\ket{0_{\rm Y}}_{\rm A}\ket{1_{\rm Y}}_{\rm B}-\ket{1_{\rm Y}}_{\rm A}\ket{0_{\rm Y}}_{\rm B}$, we
may conclude that for $t_{\rm E}=2$, we adopt the definition of
the phase error such that it is a coincidence event in Alice and Bob's fictitious Y basis measurements given the Z basis state
preparation.

\section{Security proof in the finite key size regime}\label{complete proof}
In this Appendix, we present an information theoretic security proof in the finite key size regime.

\subsection{The key length in the finite key size regime} \label{Appendix: phase-number}
The security proof in the finite key size regime is based on the fictitious protocol
we introduced in Appendix \ref{Appendix-fictitious}. In particular, we directly borrow results
and arguments made in
Appendix \ref{Appendix-fictitious}-\ref{subsection: Fair sampling}, and we start with considering an event with ${C'=0}$,
$\mu_{\rm A}=\mu_{\rm B}=\mu$ in the Code mode, and the state corresponding to this event is

\begin{eqnarray}
&&\ket{\Psi(\theta_{\rm A}, \theta_{\rm B}, \mu, \mu)}_{\rm C, A, B, E1, E2}\nonumber\\
&=&\sqrt{p_{\rm Z}^{\rm (AB)}}\ket{0_{\rm Z}}_{\rm C}\ket{\Psi_{\rm Z_A}(\theta_{\rm A}, \mu)}_{\rm A, E1}\ket{\Psi_{\rm Z_B}(\theta_{\rm B}, \mu)}_{\rm B, E2}+\sqrt{p_{\rm Y}^{\rm (AB)}}\ket{1_{\rm Z}}_{\rm C}\ket{\Psi_{\rm Y_A}(\theta_{\rm A}, \mu)}_{\rm A, E1}\ket{\Psi_{\rm Y_B}(\theta_{\rm B}, \mu)}_{\rm B, E2}\,,
\label{starting-ineq-appendix-same-basis}
\end{eqnarray} 
which can be rewritten as
\begin{eqnarray}
&&\ket{\Psi(\theta_{\rm A}, \theta_{\rm B}, \mu, \mu)}_{\rm C, A, B, E1, E2}\nonumber\\
&=&\ket{0_{\rm X}}_{\rm C}\left (p_{\rm Z}^{\rm (AB)}\ket{\Psi_{\rm Z_A}(\theta_{\rm A}, \mu)}_{\rm A, E1}\ket{\Psi_{\rm Z_B}(\theta_{\rm B}, \mu)}_{\rm B, E2}+p_{\rm Y}^{\rm (AB)}\ket{\Psi_{\rm Y_A}(\theta_{\rm A}, \mu)}_{\rm A, E1}\ket{\Psi_{\rm Y_B}(\theta_{\rm B}, \mu)}_{\rm B, E2}\right)\nonumber\\
&+&\ket{1_{\rm X}}_{\rm C}\sqrt{p_{\rm Z}^{\rm (AB)}p_{\rm Y}^{\rm (AB)}}
\left (\ket{\Psi_{\rm Z_A}(\theta_{\rm A}, \mu)}_{\rm A, E1}\ket{\Psi_{\rm Z_B}(\theta_{\rm B}, \mu)}_{\rm B, E2}-\ket{\Psi_{\rm Y_A}(\theta_{\rm A}, \mu)}_{\rm A, E1}\ket{\Psi_{\rm Y_B}(\theta_{\rm B}, \mu)}_{\rm B, E2}\right)\,.
\end{eqnarray}
Here, $p_{\rm Z}^{\rm (AB)}:=p_{{\rm Z}_{\rm A}}p_{{\rm Z}_{\rm B}}/(p_{{\rm Z}_{\rm A}}p_{{\rm Z}_{\rm B}}+p_{{\rm Y_A}}p_{{\rm Y_B}})$,
$p_{\rm Y}^{\rm (AB)}:=p_{{\rm Y_A}}p_{{\rm Y_B}}/(p_{{\rm Z}_{\rm A}}p_{{\rm Z}_{\rm B}}+p_{{\rm Y_A}}p_{{\rm Y_B}})$,
$\ket{0_{\rm X}}_{\rm C}:=\sqrt{p_{\rm Z}^{\rm (AB)}}\ket{0_{\rm Z}}_{C}+\sqrt{p_{\rm Y}^{\rm (AB)}}
\ket{1_{\rm Z}}_{C}$, and $\ket{1_{\rm X}}_{\rm C}:=\sqrt{p_{\rm Y}^{\rm (AB)}}\ket{0_{\rm Z}}_{C}-
\sqrt{p_{\rm Z}^{\rm (AB)}}\ket{1_{\rm Z}}_{C}$. Next, we consider a probability $p_{{\rm Z_C}=1}$ ($p_{{\rm X_C}=1}$) of
obtaining the bit value 1 from measuring system C with the
$\{\ket{0_{\rm Z}}_{\rm C}, \ket{1_{\rm Z}}_{\rm C}\}$ ($\{\ket{0_{\rm X}}_{\rm C}, \ket{1_{\rm X}}_{\rm C}\}$) basis.
Recalling that the length of a Bloch vector is equal to or less than 1~\cite{LP06}, we have
\begin{eqnarray}
\left(1-2p_{{\rm Z_C}=1}\right)^2+\frac{1}{\sin^2\Theta}\left[\left(1-2p_{{\rm X_C}=1}\right)-
\left(1-2p_{{\rm Z_C}=1}\right)\cos\Theta\right]^2\le1\,,
\end{eqnarray}
where $\sin\Theta=2\sqrt{p_{\rm Z}^{\rm (AB)}p_{\rm Y}^{\rm (AB)}}$ and
$\cos\Theta=p_{\rm Z}^{\rm (AB)}-p_{\rm Y}^{\rm (AB)}$ with $0\le\Theta\le\pi$. This inequality can be simplified to
\begin{eqnarray}
1-2p_{{\rm X_C}=1}\le (p_{\rm Z}^{\rm (AB)}-p_{\rm Y}^{\rm (AB)})(1-2p_{{\rm Z_C}=1})+4\sqrt{p_{\rm Z}^{\rm (AB)}p_{\rm Y}^{\rm (AB)}}\sqrt{p_{{\rm Z_C}=1}(1-p_{{\rm Z_C}=1})}\,.
\label{generalized Bloch}
\end{eqnarray}
In the analysis in~\cite{LP06}, the starting inequality for the analysis is not this inequality but
\begin{eqnarray}
1-2p_{{\rm X_C}=1}\le2\sqrt{p_{{\rm Z_C}=1}(1-p_{{\rm Z_C}=1})}\,,
\end{eqnarray}
which is a special case of Eq. (\ref{generalized Bloch}) with $p_{\rm Z}^{\rm (AB)}=p_{\rm Y}^{\rm (AB)}=1/2$.
Now, we directly employ Eq. (\ref{generalized Bloch}) in the analysis in~\cite{WTC18} (note that
the condition ``sb'' in the analysis in~\cite{WTC18} is guaranteed in our case because
we are considering Eq. (\ref{starting-ineq-appendix-same-basis}) in which Alice and Bob's state preparations
coincide) and we obtain
\begin{eqnarray}
&&p_{{\rm Z_C}}-2\frac{p_{\rm Z_C}}{p_{\rm X_C}} p_{{\rm X_C}=1, {\rm X_C}|\xi_{{\rm Code}, t_{\rm E}}^{\mu, \mu, C'=0}, \le\Delta/2}^{(i)}\nonumber\\
&\le&(p_{\rm Z}^{\rm (AB)}-p_{\rm Y}^{\rm (AB)})\left(p_{{\rm Z_C}}-2p_{{Y_{\perp}, \rm Y_A, {\rm Z_{C}}}|\xi_{{\rm Code}, t_{\rm E}}^{\mu, \mu, C'=0}, \le\Delta/2}^{(i)}\right)\nonumber\\
&+&(p_{\rm Z}^{\rm (AB)}-p_{\rm Y}^{\rm (AB)})\left(p_{{\rm Z_C}}-2p_{{Y_{||}, \rm Y_A, {\rm Z_{C}}}|\xi_{{\rm Code}, t_{\rm E}}^{\mu, \mu, C'=0}, \le\Delta/2}^{(i)}\right)\nonumber\\
&+& 4\sqrt{p_{\rm Z}^{\rm (AB)}p_{\rm Y}^{\rm (AB)}}\sqrt{p_{{Y_{\perp}, \rm Y_A, {\rm Z_{C}}}|\xi_{{\rm Code}, t_{\rm E}}^{\mu, \mu, C'=0}, \le\Delta/2}^{(i)}\,\, p_{{Y_{\perp}, \rm Z_A, {\rm Z_{C}}}|\xi_{{\rm Code}, t_{\rm E}}^{\mu, \mu, C'=0}, \le\Delta/2}^{(i)}}\nonumber\\
&+&4\sqrt{p_{\rm Z}^{\rm (AB)}p_{\rm Y}^{\rm (AB)}}\sqrt{p_{{Y_{||}, \rm Y_A, {\rm Z_{C}}}|\xi_{{\rm Code}, t_{\rm E}}^{\mu, \mu, C'=0}, \le\Delta/2}^{(i)}\,\, p_{{Y_{||}, \rm Z_A, {\rm Z_{C}}}|\xi_{{\rm Code}, t_{\rm E}}^{\mu, \mu, C'=0}, \le\Delta/2}^{(i)}}\nonumber \\
&=&2(p_{\rm Z}^{\rm (AB)}-p_{\rm Y}^{\rm (AB)})\left(p_{{\rm Z_C}}-p_{{\rm Y_A, {\rm Z_{C}}}|\xi_{{\rm Code}, t_{\rm E}}^{\mu, \mu, C'=0}, \le\Delta/2}^{(i)}\right)\nonumber\\
&+& 4\sqrt{p_{\rm Z}^{\rm (AB)}p_{\rm Y}^{\rm (AB)}}\sqrt{p_{{Y_{\perp}, \rm Y_A, {\rm Z_{C}}}|\xi_{{\rm Code}, t_{\rm E}}^{\mu, \mu, C'=0}, \le\Delta/2}^{(i)}\,\, p_{{Y_{\perp}, \rm Z_A, {\rm Z_{C}}}|\xi_{{\rm Code}, t_{\rm E}}^{\mu, \mu, C'=0}, \le\Delta/2}^{(i)}}\nonumber\\
&+&4\sqrt{p_{\rm Z}^{\rm (AB)}p_{\rm Y}^{\rm (AB)}}\sqrt{p_{{Y_{||}, \rm Y_A, {\rm Z_{C}}}|\xi_{{\rm Code}, t_{\rm E}}^{\mu, \mu, C'=0}, \le\Delta/2}^{(i)}\,\, p_{{Y_{||}, \rm Z_A, {\rm Z_{C}}}|\xi_{{\rm Code}, t_{\rm E}}^{\mu, \mu, C'=0}, \le\Delta/2}^{(i)}}
\label{Bloch sphere prob-appendix}
\end{eqnarray}

Now, we need to convert Eq. (\ref{Bloch sphere prob-appendix}) into the inequality in terms of numbers, and for this we first
take summation over $i \in \{1, 2, \cdots, N_{\xi_{{\rm Code}, t_{\rm E}}^{\mu, \mu, C'=0}, \le\Delta/2}\}$,
and then with the help of concavity of the square root function to obtain

\begin{eqnarray}
&&p_{\rm Z_C}N_{\xi_{{\rm Code}, t_{\rm E}}^{\mu, \mu, C'=0}, \le\Delta/2} - 2\frac{p_{\rm Z_C}}{p_{\rm X_C}} \sum_{i=1}^{N_{\xi_{{\rm Code}, t_{\rm E}}^{\mu, \mu, C'=0}, \le\Delta/2}}
p_{{\rm X_C}=1, {\rm X_C}|\xi_{{\rm Code}, t_{\rm E}}^{\mu, \mu, C'=0}, \le\Delta/2}^{(i)} \nonumber\\
&\le&2(p_{\rm Z}^{\rm (AB)}-p_{\rm Y}^{\rm (AB)})\left(p_{{\rm Z_C}}N_{\xi_{{\rm Code}, t_{\rm E}}^{\mu, \mu, C'=0}, \le\Delta/2}-\sum_{i=1}^{N_{\xi_{{\rm Code}, t_{\rm E}}^{\mu, \mu, C'=0}, \le\Delta/2}}p_{{\rm Y_A, {\rm Z_{C}}}|\xi_{{\rm Code}, t_{\rm E}}^{\mu, \mu, C'=0}, \le\Delta/2}^{(i)}\right)\nonumber\\
&+& 4\sqrt{p_{\rm Z}^{\rm (AB)}p_{\rm Y}^{\rm (AB)}}\sqrt{\left(\sum_{i=1}^{N_{\xi_{{\rm Code}, t_{\rm E}}^{\mu, \mu, C'=0}, \le\Delta/2}} p_{{Y_{\perp}, \rm Y_A, {\rm Z_{C}}}|\xi_{{\rm Code}, t_{\rm E}}^{\mu, \mu, C'=0}, \le\Delta/2}^{(i)}\right)\,\, \left(\sum_{i=1}^{N_{\xi_{{\rm Code}, t_{\rm E}}^{\mu, \mu, C'=0}, \le\Delta/2}}p_{{Y_{\perp}, \rm Z_A, {\rm Z_{C}}}|\xi_{{\rm Code}, t_{\rm E}}^{\mu, \mu, C'=0}, \le\Delta/2}^{(i)}\right)}\nonumber\\
&+&4\sqrt{p_{\rm Z}^{\rm (AB)}p_{\rm Y}^{\rm (AB)}}\sqrt{\left(\sum_{i=1}^{N_{\xi_{{\rm Code}, t_{\rm E}}^{\mu, \mu, C'=0}, \le\Delta/2}}p_{{Y_{||}, \rm Y_A, {\rm Z_{C}}}|\xi_{{\rm Code}, t_{\rm E}}^{\mu, \mu, C'=0}, \le\Delta/2}^{(i)}\right)\,\, \left(\sum_{i=1}^{N_{\xi_{{\rm Code}, t_{\rm E}}^{\mu, \mu, C'=0}, \le\Delta/2}} p_{{Y_{||}, \rm Z_A, {\rm Z_{C}}}|\xi_{{\rm Code}, t_{\rm E}}^{\mu, \mu, C'=0}, \le\Delta/2}^{(i)}\right)}
\label{Bloch sphere sum}
\end{eqnarray}
Then, we apply Azuma's inequality~\cite{Azuma} to the summations of probabilities, each of which is associated
to the expectation value for the corresponding event in $N_{\xi_{{\rm Code}, t_{\rm E}}^{\mu, \mu, C'=0}, \le\Delta/2}$ times of
trials (note that the number of the trials is conceptually fixed), and we have
the relationship in terms of number as

\begin{eqnarray}
&&p_{{\rm Z_C}}\underline{N}_{\xi_{{\rm Code}, t_{\rm E}}^{\mu, \mu, C'=0}, \le\Delta/2}-2\frac{p_{\rm Z_C}}{p_{\rm X_C}} \left(\overline{N}_{{\rm X_C}=1, {\rm X_C}|\xi_{{\rm Code}, t_{\rm E}}^{\mu, \mu, C'=0}}+\overline{N}_{\xi_{{\rm Code}, t_{\rm E}}^{\mu, \mu, C'=0}, \le\Delta/2}\delta_{{\rm X_C}=1, \mu}\right)\nonumber\\
&\le&2 (p_{\rm Z}^{\rm (AB)}-p_{\rm Y}^{\rm (AB)})\left[p_{{\rm Z_C}}N_{\xi_{{\rm Code}, t_{\rm E}}^{\mu, \mu, C'=0}, \le\Delta/2}-\left(N_{{\rm Y_A, {\rm Z_{C}}}|\xi_{{\rm Code}, t_{\rm E}}^{\mu, \mu, C'=0}, \le\Delta/2}+\tilde{N}_{\xi_{{\rm Code}, t_{\rm E}}^{\mu, \mu, C'=0}, \le\Delta/2}\delta_{{\rm Y_A, Y_{\perp}}, \mu}\right)\right]\nonumber\\
&+& 4\sqrt{p_{\rm Z}^{\rm (AB)}p_{\rm Y}^{\rm (AB)}}\sqrt{\left(N_{{Y_{\perp}, \rm Y_A, {\rm Z_{C}}}|\xi_{{\rm Code}, t_{\rm E}}^{\mu, \mu, C'=0}, \le\Delta/2}+\overline{N}_{\xi_{{\rm Code}, t_{\rm E}}^{\mu, \mu, C'=0}, \le\Delta/2}\delta_{{\rm Y_A, Y_{\perp}}, \mu}\right)}\nonumber\\
&\times&\sqrt{\left(N_{{Y_{\perp}, \rm Z_A, {\rm Z_{C}}}|\xi_{{\rm Code}, t_{\rm E}}^{\mu, \mu, C'=0}, \le\Delta/2}+
\overline{N}_{\xi_{{\rm Code}, t_{\rm E}}^{\mu, \mu, C'=0}, \le\Delta/2}\delta_{{\rm Z_A, Y_{\perp}}, \mu}\right)}\nonumber\\
&+& 4\sqrt{p_{\rm Z}^{\rm (AB)}p_{\rm Y}^{\rm (AB)}}\sqrt{\left(N_{{Y_{||}, \rm Y_A, {\rm Z_{C}}}|\xi_{{\rm Code}, t_{\rm E}}^{\mu, \mu, C'=0}, \le\Delta/2}+\overline{N}_{\xi_{{\rm Code}, t_{\rm E}}^{\mu, \mu, C'=0}, \le\Delta/2}\delta_{{\rm Y_A, Y_{||}}, \mu}\right)}\nonumber\\
&\times&\sqrt{\left(N_{{Y_{||}, \rm Z_A, {\rm Z_{C}}}|\xi_{{\rm Code}, t_{\rm E}}^{\mu, \mu, C'=0}, \le\Delta/2}
+\overline{N}_{\xi_{{\rm Code}, t_{\rm E}}^{\mu, \mu, C'=0}, \le\Delta/2}\delta_{{\rm Z_A, Y_{||}}, \mu}\right)}\,,
\label{azuma-number-appendix}
\end{eqnarray}
which holds probability at least $1-\epsilon_{{\rm X_C}=1, \mu}-\epsilon_{{\rm Y_A, Y_{\perp}}, \mu}-\epsilon_{{\rm Z_A, Y_{\perp}}, \mu}-\epsilon_{{\rm Y_A, Y_{||}}, \mu}-\epsilon_{{\rm Z_A, Y_{||}}, \mu}$ (see Appendix \ref{summary} for the relationships between
$\epsilon$'s and $\delta$'s, in which $\epsilon$'s are any positive value). 
Here, each $\epsilon$ represents a failure probability of each of the estimation, and 
$\tilde{N}_{\xi_{{\rm Code}, t_{\rm E}}^{\mu, \mu, C'=0}, \le\Delta/2}=\overline{N}_{\xi_{{\rm Code}, t_{\rm E}}^{\mu, \mu, C'=0}, \le\Delta/2}$ if $p_{\rm Z}^{\rm (AB)}\le p_{\rm Y}^{\rm (AB)}$, and $\tilde{N}_{\xi_{{\rm Code}, t_{\rm E}}^{\mu, \mu, C'=0}, \le\Delta/2}=\underline{N}_{\xi_{{\rm Code}, t_{\rm E}}^{\mu, \mu, C'=0}, \le\Delta/2}$ if $p_{\rm Z}^{\rm (AB)}\ge p_{\rm Y}^{\rm (AB)}$, and
we used
$N_{{\rm X_C}=1, {\rm X_C}|\xi_{{\rm Code},
t_{\rm E}}^{\mu, \mu, C'=0}}\ge N_{{\rm X_C}=1, {\rm X_C}|\xi_{{\rm Code},
t_{\rm E}}^{\mu, \mu, C'=0}, \le\Delta/2}$ where $N_{{\rm X_C}=1, {\rm X_C}|\xi_{{\rm Code},
t_{\rm E}}^{\mu, \mu, C'=0}, \le\Delta/2}$ is the number of the events ${\rm X_C}=1$ and ${\rm X_C}$
among the events specified by $\xi_{{\rm Code}, t_{\rm E}}^{\mu, \mu, C'=0}$ and $\le\Delta/2$. Note that
$\underline{N}_{\xi_{{\rm Code}, t_{\rm E}}^{\mu, \mu, C'=0}, \le\Delta/2}$ and $\overline{N}_{\xi_{{\rm Code}, t_{\rm E}}^{\mu, \mu, C'=0}, \le\Delta/2}$ are not directly obtained in the
experiment. This is so because we have a decomposition
$N_{\xi_{{\rm Code}, t_{\rm E}}^{\mu, \mu, C'=0}, \le\Delta/2}=N_{{\rm Z_C}|\xi_{{\rm Code}, t_{\rm E}}^{\mu, \mu, C'=0}, \le\Delta/2}+N_{{\rm X_C}|\xi_{{\rm Code}, t_{\rm E}}^{\mu, \mu, C'=0}, \le\Delta/2}$, and
$N_{{\rm X_C}|\xi_{{\rm Code}, t_{\rm E}}^{\mu, \mu, C'=0}, \le\Delta/2}$ cannot be directly obtained in the actual
protocol (Alice and Bob do not announce their basis selections when the ${\rm X_C}$ basis is chosen in the Code mode).
See Appendix \ref{Appendix-lower-upper-code} for the derivation of the bounds and the explicit forms,
which holds probability at least $1-\underline{\epsilon}_{{\rm C}, \mu, \le\Delta/2}-\overline{\epsilon}_{{\rm C}, \mu, \le\Delta/2}$.
By taking the asymptotic limit of, i.e. neglecting
$\delta$'s and the bounds of the numbers, we have the inequality presented in
Eq. (\ref{azuma-number-asympt}).

As for the key length, given the upper bound of the number of phase errors, the key length $l$ is expressed as~\cite{key length, SCIC}

\begin{eqnarray}
l_{\mu, t_{\rm E}=1}&=&N_{{\rm Z_A, {\rm Z_{C}}}|\xi_{{\rm Code}, t_{\rm E}=1}^{\mu, \mu, C'=0}, \le\Delta/2}\left[1-
h\left(\frac{\overline{N}_{{Y_{||}, \rm Z_A, {\rm Z_{C}}}|\xi_{{\rm Code}, t_{\rm E}=1}^{\mu, \mu, C'=0}, \le\Delta/2}}{N_{{\rm Z_A, {\rm Z_{C}}}|\xi_{{\rm Code}, t_{\rm E}=1}^{\mu, \mu, C'=0}, \le\Delta/2}}\right)\right]-\log_{2}\frac{2}{\epsilon_{{\rm PA}, \mu}}-\lambda_{{\rm EC}, \mu}\,,\\
l_{\mu, t_{\rm E}=2}&=&N_{{\rm Z_A, {\rm Z_{C}}}|\xi_{{\rm Code}, t_{\rm E}=2}^{\mu, \mu, C'=0}, \le\Delta/2}\left[1-
h\left(\frac{\overline{N}_{{Y_{\perp}, \rm Z_A, {\rm Z_{C}}}|\xi_{{\rm Code}, t_{\rm E}=2}^{\mu, \mu, C'=0}, \le\Delta/2}}{N_{{\rm Z_A, {\rm Z_{C}}}|\xi_{{\rm Code}, t_{\rm E}=2}^{\mu, \mu, C'=0}, \le\Delta/2}}\right)\right]-\log_{2}\frac{2}{\epsilon_{{\rm PA}, \mu}}-\lambda_{{\rm EC}, \mu} \,,
\label{key length}
\end{eqnarray}
where, $h(x)$ is the binary entropy function, and $\lambda_{{\rm EC}, \mu}$ is the amount of information exchanged for error correction.
Here, when we define $\epsilon_{{\rm PE}, \mu}$ as the probability that the phase error estimation fails and choose a $\epsilon_{{\rm PA}, \mu}$,
then the key is $\epsilon_{{\rm s}, \mu}$-secret with $\epsilon_{{\rm s}, \mu}:=\sqrt{2}\sqrt{\epsilon_{{\rm PA}, \mu}+\epsilon_{{\rm PE}, \mu}}$,
where

\begin{eqnarray}
\epsilon_{{\rm PE}, \mu}:=\epsilon_{{\rm X_C}=1, \mu}+\epsilon_{{\rm Y_A, Y_{\perp}}, \mu}+\epsilon_{{\rm Z_A, Y_{\perp}}, \mu}+\epsilon_{{\rm Y_A, Y_{||}}, \mu}+\epsilon_{{\rm Z_A, Y_{||}}, \mu}+\underline{\epsilon}_{{\rm C}, \mu, \le\Delta/2}+\overline{\epsilon}_{{\rm C}, \mu, \le\Delta/2}+\epsilon_{{\rm X_C}=1, {\rm est}, \mu}\,,
\label{key: epsilon}
\end{eqnarray}
where $\epsilon_{{\rm X_C}=1, {\rm est}, \mu}$ is the failure probability of estimating
${\overline N}_{{\rm X_C}=1, {\rm X_C}|\xi_{{\rm Code}, t_{\rm E}}^{\mu, \mu, C'=0}}$, which will be given by Eq. (\ref{final epsilon}).
From next subsections, we derive $\overline{N}_{{\rm X_C}=1, {\rm X_C}|\xi_{{\rm Code}, t_{\rm E}}^{\mu, \mu, C'=0}}$.

\subsection{Estimation of $N_{{\rm X_C}=1, {\rm X_C}|\xi_{{\rm Code}, t_{\rm E}}^{\mu, \mu, C'=0}}$ from
$N_{{\rm X_C}=1, {\rm X_C}|\xi_{{\rm Test}, t_{\rm E}}^{\mu, \mu, C'=0}}$}

In this subsection, we explain the estimation of ${\overline N}_{{\rm X_C}=1, {\rm X_C}|\xi_{{\rm Code}, t_{\rm E}}^{\mu, \mu, C'=0}}$, which is an upper bound of $N_{{\rm X_C}=1, {\rm X_C}|\xi_{{\rm Code}, t_{\rm E}}^{\mu, \mu, C'=0}}$, from $N_{{\rm X_C}=1, {\rm X_C}|\xi_{{\rm Test}, t_{\rm E}}^{\mu, \mu, C'=0}}$.
First, recall the discussion in Sec. \ref{subsection: Fair sampling} that the choice between the Code and the Test modes
for ${\rm Z_{C'}}=0$ and $\mu_{\rm A}=\mu_{\rm A}=\mu$ is independent of any other events, and we employ this argument to estimate
$\overline{N}_{{\rm X_C}=1, {\rm X_C}|\xi_{{\rm Code}, t_{\rm E}}^{\mu, \mu, C'=0}}$
from $N_{{\rm X_C}=1, {\rm X_C}|\xi_{{\rm Test}, t_{\rm E}}^{\mu, \mu, C'=0}}$. Next, observe that $N_{{\rm X_C}=1, {\rm X_C}|\xi_{{\rm Code}, t_{\rm E}}^{\mu, \mu, C'=0}}$
and $N_{{\rm X_C}=1, {\rm X_C}|\xi_{{\rm Test}, t_{\rm E}}^{\mu, \mu, C'=0}}$ remain unchanged even if we perform the ${\rm X_{C}}$ basis measurement
on systems C in the Code mode with the selection of ${\rm Z_C}$ basis. This is so because measurements
on different systems commute. Therefore, only for the purpose for
estimating $N_{{\rm X_C}=1, {\rm X_C}|\xi_{{\rm Code}, t_{\rm E}}^{\mu, \mu, C'=0}}$
from $N_{{\rm X_C}=1, {\rm X_C}|\xi_{{\rm Test}, t_{\rm E}}^{\mu, \mu, C'=0}}$, we are allowed to suppose that
Alice measures systems C with the ${\rm X_{C}}$ basis, and then each of the instances with ${\rm X_{C}}=1$ is assigned either
to the Test mode or to the selection of ${\rm X_C}$ basis in the Code mode with probabilities $s_{{\rm X}_{\rm C}}$ and $1-s_{{\rm X}_{\rm C}}$, respectively.
Here, $s_{{\rm X}_{\rm C}}:=p_{\rm T}/(p_{\rm T}+p_{\rm C}p_{{\rm X}_{\rm C}})$.
That is, we have $N_{{\rm X_C}=1, {\rm X_C}|\xi_{{\rm Code}, t_{\rm E}}^{\mu, \mu, C'=0}}+N_{{\rm X_C}=1, {\rm X_C}|\xi_{{\rm Test}, t_{\rm E}}^{\mu, \mu, C'=0}}$ of 1's and
these 1's are assigned either to the Test or Code modes with the Bernoulli trials. Moreover,
by recalling that the more event ${\rm X_C}=1$ we have the more information leakage occurs, we consider a pessimistic situation that we have $N_{{\rm X_C}=1, {\rm X_C}|\xi_{{\rm Code}, t_{\rm E}}^{\mu, \mu, C'=0}}+{\overline N}_{{\rm X_C}=1, {\rm X_C}|\xi_{{\rm Test}, t_{\rm E}}^{\mu, \mu, C'=0}}$ of 1's in total.
Noting that the number of the trials is conceptually fixed, and this trial is an identical and independent trial, we can use the Chernoff bound \cite{Chernoff}, and we have that
\begin{eqnarray}
N_{{\rm X_C}=1, {\rm X_C}|\xi_{{\rm Code}, t_{\rm E}}^{\mu, \mu, C'=0}}\le \frac{1-s_{{\rm X}_{\rm C}}+\delta_{{\rm TC}, \mu}}{s_{{\rm X}_{\rm C}}-\delta_{{\rm TC}, \mu}}\overline{N}_{{\rm X_C}=1, {\rm X_C}|\xi_{{\rm Test}, t_{\rm E}}^{\mu, \mu, C'=0}}\nonumber\\
\label{estimate-code-test}
\end{eqnarray}
holds with probability at least $1-\epsilon_{{\rm TC}, \mu}$ (see Appendix \ref{summary} for the relationship between
$\epsilon_{{\rm TC}, \mu}$ and $\delta_{{\rm TC}, \mu}$).
Next problem is to estimate $\overline{N}_{{\rm X_C}=1, {\rm X_C}|\xi_{{\rm Test}, t_{\rm E}}^{\mu,
\mu, C'=0}}$ by using the decoy state method, which we present in the next section. 

\subsection{Estimation of $\overline{N}_{{\rm X_C}=1, {\rm X_C}|\xi_{{\rm Test}, t_{\rm E}}^{\mu,
\mu, C'=0}}$ and $\underline{N}_{{\rm X_C}=1, {\rm X_C}|\xi_{{\rm Test}, t_{\rm E}}^{\mu,
\mu, C'=0}}$ using the decoy state method}\label{subsection-decoy}

In this section, we present how to estimate $\overline{N}_{{\rm X_C}=1, {\rm X_C}|\xi_{{\rm Test}, t_{\rm E}}^{\mu,
\mu, C'=0}}$.
For this, we start with Eq. (\ref{first eq-main}), which means that our problem is reduced to the estimation of $N_{{\rm X_C}|\xi_{{\rm Test}, t_{\rm E}}^{\mu, \mu, C'=0}, n_{\rm A}, n_{\rm B}}$
 for various $(n_{\rm A}, n_{\rm B})$.
Next, recall the standard decoy state argument that when Alice and Bob respectively emit $n_{\rm A}$ and $n_{\rm B}$ photons
to Charlie, those photons do not contain any information about the intensity setting. This is so because we assume that there is no
state preparation flaw and side channel. Therefore, one can imagine that
Alice and Bob perform the photon number measurements first, and then they probabilistically assign their intensity settings
{\it after} Charlie announces his detection result $t_{\rm E}$. With this observation, we first define the following
expected quantities for each of the combinations of the intensity settings as
\begin{eqnarray}
{\rm Ex}_{\mu_{\rm A}, \mu_{\rm B}|\xi_{{\rm Test}, t_{\rm E}}^{C'=0}}:=
\sum_{n_{\rm A}, n_{\rm B}} N_{n_{\rm A}, n_{\rm B}|\xi_{{\rm Test}, t_{\rm E}}^{C'=0}}q_{\mu_{\rm A}, \mu_{\rm B}|n_{\rm A}, n_{\rm B}}\,.\nonumber\\
\label{decoy ineq 1}
\end{eqnarray}
Here, $N_{n_{\rm A}, n_{\rm B}|\xi_{{\rm Test}, t_{\rm E}}^{C'=0}}$ is the number of $n_{\rm A}$ and $n_{\rm B}$ photon emission
events among events $\xi_{{\rm Test}, t_{\rm E}}^{C'=0}$, and
$q_{\mu_{\rm A}, \mu_{\rm B}|n_{\rm A}, n_{\rm B}}$
is a probability that Alice and Bob respectively select an intensity setting $\mu_{\rm A}$ and $\mu_{\rm B}$,
given that Alice and Bob respectively emit $n_{\rm A}$ and $n_{\rm B}$ photons (the explicit form of
$q_{\mu_{\rm A}, \mu_{\rm B}|n_{\rm A}, n_{\rm B}}$ is given in Appendix \ref{explicit-form}).
By noting that these expectation values are associated to independent but non-identical trials
whose number is conceptually fixed, we can apply the Hoeffding's inequality~\cite{Hoeffding} to them, and we obtain
\begin{eqnarray}
\left| N_{{\rm X_C}|\xi_{{\rm Test}, t_{\rm E}}^{\mu_{\rm A}, \mu_{\rm B}, C'=0}}-{\rm Ex}_{\mu_{\rm A}, \mu_{\rm B}|\xi_{{\rm Test}, t_{\rm E}}^{C'=0}}\right|&\le& N_{{\rm X_C}|\xi_{{\rm Test}, t_{\rm E}}^{\mu_{\rm A}, \mu_{\rm B}, C'=0}}\delta_{\mu_{\rm A}, \mu_{\rm B}}\,,\label{hof3}
\end{eqnarray}
which holds probability at least $1-2\epsilon_{\mu_{\rm A}, \mu_{\rm B}}$ (see Appendix \ref{summary} for the relationship
between $\epsilon_{\mu_{\rm A}, \mu_{\rm B}}$ and $\delta_{\mu_{\rm A}, \mu_{\rm B}}$). Here, note that
$N_{{\rm X_C}|\xi_{{\rm Test}, t_{\rm E}}^{\mu_{\rm A}, \mu_{\rm B}, C'=0}}$ is available in the actual protocol because
Alice and Bob exchange the bases information in the event $\xi_{{\rm Test}, t_{\rm E}}^{\mu_{\rm A}, \mu_{\rm B}, C'=0}$ (recall
that ${\rm X_C}|$ represents the Gedanken measurement, and the ${\rm Z_C}$ basis is used in the fictitious protocol).

From Eqs. (\ref{decoy ineq 1})-(\ref{hof3}), we can numerically obtain a lower and a upper bounds of
$N_{n_{\rm A}, n_{\rm B}|\xi_{{\rm Test}, t_{\rm E}}^{C'=0}}$ using the experimentally available data, and we denote them by
$\underline{N}_{n_{\rm A}, n_{\rm B}|\xi_{{\rm Test}, t_{\rm E}}^{C'=0}}$ and
$\overline{N}_{n_{\rm A}, n_{\rm B}|\xi_{{\rm Test}, t_{\rm E}}^{C'=0}}$, which is valid probability at least $1-\epsilon_{\rm decoy, Fock}$ with
\begin{eqnarray}
\epsilon_{\rm decoy, Fock}:=\sum_{\mu_{\rm A}, \mu_{\rm B}}2\epsilon_{\mu_{\rm A}, \mu_{\rm B}}.
\end{eqnarray}
After obtaining the lower bound, we consider to probabilistically assign intensity settings to those photon number instances.
For this, we consider the following expectation values

\begin{eqnarray}
\sum_{n_{\rm A}, n_{\rm B}|(n_{\rm A}, n_{\rm B})\in \{(0,0), (1,0), (0,1), (1,1)\}} \underline{N}_{n_{\rm A}, n_{\rm B}|\xi_{{\rm Test}, t_{\rm E}}^{C'=0}}q_{\mu, \mu|n_{\rm A}, n_{\rm B}}\,,
\label{decoy ineq 2}
\end{eqnarray}
which can be associated to the actual number by using the Hoeffding's inequality to have that
\begin{eqnarray}
\underline{N}_{{\rm X_C}|\xi_{{\rm Test}, t_{\rm E}}^{\mu, \mu, C'=0}, n_{\rm A}\le1, n_{\rm B}\le1}&=&
\sum_{n_{\rm A}, n_{\rm B}|(n_{\rm A}, n_{\rm B})\in \{(0,0), (1,0), (0,1), (1,1)\}} \underline{N}_{n_{\rm A}, n_{\rm B}|\xi_{{\rm Test}, t_{\rm E}}^{C'=0}}q_{\mu, \mu|n_{\rm A}, n_{\rm B}}\nonumber\\
&-&\left(\sum_{n_{\rm A}, n_{\rm B}|(n_{\rm A}, n_{\rm B})\in \{(0,0), (1,0), (0,1), (1,1)\}} \overline{N}_{n_{\rm A}, n_{\rm B}|\xi_{{\rm Test}, t_{\rm E}}^{C'=0}}\right)\delta_{\mu, \mu|n_{\rm A}\le1, n_{\rm B}\le1}\nonumber\label{lower-1-1-better}\,\\
\end{eqnarray}
holds probability at least $1-\epsilon_{\mu, \mu|n_{\rm A}\le1, n_{\rm B}\le1}$ (see Appendix \ref{summary} for
the relationship between $\epsilon_{\mu, \mu|n_{\rm A}\le1, n_{\rm B}\le1}$ and
$\delta_{\mu, \mu|n_{\rm A}\le1, n_{\rm B}\le1}$). From this, we have
\begin{eqnarray}
\overline{N}_{{\rm X_C}|\xi_{{\rm Test}, t_{\rm E}}^{\mu, \mu, C'=0}}=N_{{\rm X_C}|\xi_{{\rm Test}, t_{\rm E}}^{\mu, \mu, C'=0}}-\underline{N}_{{\rm X_C}|\xi_{{\rm Test}, t_{\rm E}}^{\mu, \mu, C'=0}, n_{\rm A}\le1, n_{\rm B}\le1}\,,
\label{upper-Xc=1-test}
\end{eqnarray}
leading to 
\begin{eqnarray}
\overline{N}_{{\rm X_C}=1, {\rm X_C}|\xi_{{\rm Code}, t_{\rm E}}^{\mu, \mu, C'=0}}&:=&\frac{1-s_{{\rm X}_{\rm C}}+\delta_{{\rm TC}, \mu}}{s_{{\rm X}_{\rm C}}-\delta_{{\rm TC}, \mu}} \Big(N_{{\rm X_C}|\xi_{{\rm Test}, t_{\rm E}}^{\mu, \mu, C'=0}}-\underline{N}_{{\rm X_C}|\xi_{{\rm Test}, t_{\rm E}}^{\mu, \mu, C'=0}, n_{\rm A}\le1, n_{\rm B}\le1}\Big)\,.
\end{eqnarray}

Finally, by taking a summation over all $\epsilon$'s appearing in this subsection, we have 
the failure probability of estimating
${\overline N}_{{\rm X_C}=1, {\rm X_C}|\xi_{{\rm Code}, t_{\rm E}}^{\mu, \mu, C'=0}}$ as 
\begin{eqnarray}
\epsilon_{{\rm X_C}=1, {\rm est}, \mu}&=&\epsilon_{{\rm TC}, \mu}+\epsilon_{\rm decoy, Fock}+
\epsilon_{\mu, \mu|n_{\rm A}\le1, n_{\rm B}\le1}\,.
\label{final epsilon}
\end{eqnarray}

\section{Bounds of  $N_{{\rm X_C}|\xi_{{\rm Code}, t_{\rm E}}^{\mu, \mu, C'=0}, \le\Delta/2}$}\label{Appendix-lower-upper-code}

In this Appendix, we estimate bounds of $N_{{\rm X_C}|\xi_{{\rm Code}, t_{\rm E}}^{\mu, \mu, C'=0}, \le\Delta/2}$.
For the estimation, we exploit the fact that the ${\rm Z}_{\rm C}$ basis or the ${\rm X}_{\rm C}$ basis is chosen probabilistically in the Code mode. In this case, $N_{{\rm X_C}|\xi_{{\rm Code}, t_{\rm E}}^{\mu, \mu, C'=0}, \le\Delta/2}$ ($N_{{\rm Z_C}|\xi_{{\rm Code}, t_{\rm E}}^{\mu, \mu, C'=0}, \le\Delta/2}$)
is an unknown (a known) quantity in the actual protocol. Then, we imagine that we conduct
$N_{{\rm X_C}|\xi_{{\rm Code}, t_{\rm E}}^{\mu, \mu, C'=0}, \le\Delta/2}+N_{{\rm Z_C}|\xi_{{\rm Code}, t_{\rm E}}^{\mu, \mu, C'=0}, \le\Delta/2}$ times of the Bernoulli trials (note that this number is conceptually fixed), in which the ${\rm Z}_{\rm C}$ basis and
the ${\rm X}_{\rm C}$ basis are selected with probability $p_{{\rm Z}_{\rm C}}$ and $p_{{\rm X}_{\rm C}}$, respectively. Thanks to the fact that this trial is an identical and independent trial, we can use the Chernoff bound, we have for each $\mu\in \{\mu_{1}, \mu_{2}, \mu_3\}$ that
\begin{eqnarray}
\underline{N}_{{\rm X_C}|\xi_{{\rm Code}, t_{\rm E}}^{\mu, \mu, C'=0}, \le\Delta/2}&\le& N_{{\rm X_C}|\xi_{{\rm Code}, t_{\rm E}}^{\mu, \mu, C'=0}, \le\Delta/2}\le \overline{N}_{{\rm X_C}|\xi_{{\rm Code}, t_{\rm E}}^{\mu, \mu, C'=0}, \le\Delta/2}\,,\label{ineq0}
\end{eqnarray}
where
\begin{eqnarray}
\underline{N}_{{\rm X_C}|\xi_{{\rm Code}, t_{\rm E}}^{\mu, \mu, C'=0}, \le\Delta/2}&:=&\frac{1-p_{{\rm Z}_{\rm C}}-\underline{\delta}_{{\rm C}, \mu, \le\Delta/2}}{p_{{\rm Z}_{\rm C}}+\underline{\delta}_{{\rm C}, \mu, \le\Delta/2}}N_{{\rm Z_C}|\xi_{{\rm Code}, t_{\rm E}}^{\mu, \mu, C'=0, \le\Delta/2}}\nonumber\\
\overline{N}_{{\rm X_C}|\xi_{{\rm Code}, t_{\rm E}}^{\mu, \mu, C'=0}, \le\Delta/2}&:=&
\frac{1-p_{{\rm Z}_{\rm C}}+\overline{\delta}_{{\rm C}, \mu, \le\Delta/2}}{p_{{\rm Z}_{\rm C}}-\overline{\delta}_{{\rm C}, \mu, \le\Delta/2}}N_{{\rm Z_C}|\xi_{{\rm Code}, t_{\rm E}}^{\mu, \mu, C'=0}, \le\Delta/2}\nonumber\\
\end{eqnarray}
holds at least probability $1-\underline{\epsilon}_{{\rm C}, \mu, \le\Delta/2}-\overline{\epsilon}_{{\rm C}, \mu, \le\Delta/2}$ with $\underline{\epsilon}_{{\rm C}, \mu, \le\Delta/2}:=e^{-D(p_{{\rm Z}_{\rm C}}+\underline{\delta}_{{\rm C}, \mu, \le\Delta/2}||p_{{\rm Z}_{\rm C}})
N_{{\rm Z_C}|\xi_{{\rm Code}, t_{\rm E}}^{\mu, \mu, C'=0}}, \le\Delta/2}$ and
$\overline{\epsilon}_{{\rm C}, \mu, \le\Delta/2}:=e^{-D(p_{{\rm X}_{\rm C}}-\overline{\delta}_{{\rm C}, \mu, \le\Delta/2}||p_{{\rm X}_{\rm C}})
N_{{\rm Z_C}|\xi_{{\rm Code}, t_{\rm E}}^{\mu, \mu, C'=0}}, \le\Delta/2}$.

Importantly, notice that
the upper and lower bounds are expressed by $N_{{\rm Z_C}|\xi_{{\rm Code}, t_{\rm E}}^{\mu, \mu, C'=0}, \le\Delta/2}$, which is
available in the actual protocol. With these bounds, we have
\begin{eqnarray}
\underline{N}_{\xi_{{\rm Code}, t_{\rm E}}^{\mu, \mu, C'=0}, \le\Delta/2}&:=&N_{{\rm Z_C}|\xi_{{\rm Code}, t_{\rm E}}^{\mu, \mu, C'=0}, \le\Delta/2}+\underline{N}_{{\rm X_C}|\xi_{{\rm Code}, t_{\rm E}}^{\mu, \mu, C'=0}, \le\Delta/2}\,,\label{lower-total-number}\\
\overline{N}_{\xi_{{\rm Code}, t_{\rm E}}^{\mu, \mu, C'=0}, \le\Delta/2}&:=&N_{{\rm Z_C}|\xi_{{\rm Code}, t_{\rm E}}^{\mu, \mu, C'=0}, \le\Delta/2}+\overline{N}_{{\rm X_C}|\xi_{{\rm Code}, t_{\rm E}}^{\mu, \mu, C'=0}, \le\Delta/2}\,,
\end{eqnarray}
which are used in Eq. (\ref{azuma-number-appendix}).

\section{Summary of the relationships between $\epsilon$'s and $\delta$'s}\label{summary}
In this section, we summarize all the relationships between $\epsilon$'s and $\delta$'s. For this,
we define $f_{\rm Az}(x,y):=\sqrt{(2/x)\ln(1/y)}$, $D(x||y):=x\ln\frac{x}{y}+(1-x)\ln\left(\frac{1-x}{1-y}\right)$,
and $f_{\rm Hoe}(x,y):=\sqrt{1/(2x)\ln (1/y)}$. With these definitions, the relationships are given as follows:

\begin{enumerate}

\item $\delta_{{\rm X_C}=1, \mu}=f_{\rm Az}(\underline{N}_{\xi_{{\rm Code}, t_{\rm E}}^{\mu, \mu, C'=0}, \le\Delta/2}, \epsilon_{{\rm X_C}=1, \mu})$. See Eq. (\ref{lower-total-number}) for the definition of $\underline{N}_{\xi_{{\rm Code}, t_{\rm E}}^{\mu, \mu, C'=0}, \le\Delta/2}$.

\item $\delta_{{\rm Y_A, Y_{\perp}}, \mu}=f_{\rm Az}(\underline{N}_{\xi_{{\rm Code}, t_{\rm E}}^{\mu, \mu, C'=0}, \le\Delta/2}, \epsilon_{{\rm Y_A, Y_{\perp}}, \mu})$. See Eq. (\ref{lower-total-number}) for the definition of $\underline{N}_{\xi_{{\rm Code}, t_{\rm E}}^{\mu, \mu, C'=0}, \le\Delta/2}$.

\item $\delta_{{\rm Z_A, Y_{\perp}}, \mu}=f_{\rm Az}(\underline{N}_{\xi_{{\rm Code}, t_{\rm E}}^{\mu, \mu, C'=0}, \le\Delta/2}, \epsilon_{{\rm Z_A, Y_{\perp}}, \mu})$. See Eq. (\ref{lower-total-number}) for the definition of $\underline{N}_{\xi_{{\rm Code}, t_{\rm E}}^{\mu, \mu, C'=0}, \le\Delta/2}$.

\item $\delta_{{\rm Y_A, Y_{||}}, \mu}=f_{\rm Az}(\underline{N}_{\xi_{{\rm Code}, t_{\rm E}}^{\mu, \mu, C'=0}, \le\Delta/2}, \epsilon_{{\rm Y_A, Y_{||}}, \mu})$. See Eq. (\ref{lower-total-number}) for the definition of $\underline{N}_{\xi_{{\rm Code}, t_{\rm E}}^{\mu, \mu, C'=0}, \le\Delta/2}$.

\item $\delta_{{\rm Z_A, Y_{||}}, \mu}=f_{\rm Az}(\underline{N}_{\xi_{{\rm Code}, t_{\rm E}}^{\mu, \mu, C'=0}, \le\Delta/2}, \epsilon_{{\rm Z_A, Y_{||}}, \mu})$. See Eq. (\ref{lower-total-number}) for the definition of $\underline{N}_{\xi_{{\rm Code}, t_{\rm E}}^{\mu, \mu, C'=0}, \le\Delta/2}$.

\item $\underline{\epsilon}_{{\rm C}, \mu, \le\Delta/2}:=e^{-D(p_{{\rm Z}_{\rm C}}+\underline{\delta}_{{\rm C}, \mu, \le\Delta/2}||p_{{\rm Z}_{\rm C}})
N_{{\rm Z_C}|\xi_{{\rm Code}, t_{\rm E}}^{\mu, \mu, C'=0}}, \le\Delta/2}$ and
$\overline{\epsilon}_{{\rm C}, \mu, \le\Delta/2}:=e^{-D(p_{{\rm X}_{\rm C}}-\overline{\delta}_{{\rm C}, \mu, \le\Delta/2}||p_{{\rm X}_{\rm C}})
N_{{\rm Z_C}|\xi_{{\rm Code}, t_{\rm E}}^{\mu, \mu, C'=0}}, \le\Delta/2}$.

\item $\epsilon_{{\rm TC}, \mu}:=e^{-D((1-s_{{\rm X}_{\rm C}})+
\delta_{{\rm TC}, \mu}||(1-s_{{\rm X}_{\rm C}}))\overline{N}_{{\rm X_C}=1, {\rm X_C}|\xi_{{\rm Test}, t_{\rm E}}^{\mu, \mu, C'=0}}}$.
Here, $s_{{\rm X}_{\rm C}}:=p_{\rm T}/(p_{\rm T}+p_{\rm C}p_{\rm X_C})$.

\item $\delta_{\mu_{\rm A}, \mu_{\rm B}}=f_{\rm Hoe}(N_{{\rm X_C}|\xi_{{\rm Test}, t_{\rm E}}^{\mu_{\rm A}, \mu_{\rm B}, C'=0}}, \epsilon_{\mu_{\rm A}, \mu_{\rm B}|n_{\rm A}, n_{\rm B}})$. Note that we have 9
$\delta_{\mu_{\rm A}, \mu_{\rm B}}$'s because we have 9 combinations of Alice and Bob's intensity settings.


\item 
\begin{eqnarray*}
\delta_{\mu, \mu|n_{\rm A}\le1, n_{\rm B}\le1}=f_{\rm Hoe}\left(\sum_{n_{\rm A}, n_{\rm B}|(n_{\rm A}, n_{\rm B})\in \{(0,0), (1,0), (0,1), (1,1)\}} \underline{N}_{n_{\rm A}, n_{\rm B}|\xi_{{\rm Test}, t_{\rm E}}^{C'=0}}, \epsilon_{\mu, \mu|n_{\rm A}\le1, n_{\rm B}\le1}\right)\,.
\end{eqnarray*}
Here, $\underline{N}_{n_{\rm A}, n_{\rm B}|\xi_{{\rm Test}, t_{\rm E}}^{C'=0}}$ is obtained by the decoy state
method.

\end{enumerate}

\section{Probability
of obtaining ${\rm X_C}=1$ for $(n_{\rm A}, n_{\rm B})\in \{(0,0), (1,0), (0,1), (1,1)\}$}\label{zero-pro}

In this appendix, we show that the probability
of obtaining ${\rm X_C}=1$ for $(n_{\rm A}, n_{\rm B})\in \{(0,0), (1,0), (0,1), (1,1)\}$ is zero. For this, recall that we have
the definitions of the states as

\begin{eqnarray}
\ket{\Psi_{\rm Z_A}(\theta_{\rm A}, \mu_{\rm A})}_{\rm A, E1}&:=&\frac{1}{\sqrt{2}}\left(\ket{0_{\rm Z}}_{\rm A}(\ket{e^{i \theta_{\rm A}}\sqrt{\mu_{\rm A}}}_{\rm ref}\ket{e^{i \theta_{\rm A}}\sqrt{\mu_{\rm A}}}_{\rm sg})_{\rm E1}+\ket{1_{\rm Z}}_{\rm A}(\ket{e^{i \theta_{\rm A}}\sqrt{\mu_{\rm A}}}_{\rm ref}\ket{e^{i (\theta_{\rm A}+\pi)}\sqrt{\mu_{\rm A}}}_{\rm sg})_{\rm E1}\right),\,\nonumber\\
\ket{\Psi_{\rm Y_A}(\theta_{\rm A}, \mu_{\rm A})}_{\rm A, E1}&:=&\frac{1}{\sqrt{2}}\left(\ket{1_{\rm Y}}_{\rm A}(\ket{e^{i \theta_{\rm A}}\sqrt{\mu_{\rm A}}}_{\rm ref}\ket{e^{i (\theta_{\rm A}+\pi/2)}\sqrt{\mu_{\rm A}}}_{\rm sg})_{\rm E1}+\ket{0_{\rm Y}}_{\rm A}(\ket{\sqrt{e^{i \theta_{\rm A}}\mu_{\rm A}}}_{\rm ref}\ket{e^{i (\theta_{\rm A}+3\pi/2)}\sqrt{\mu_{\rm A}}}_{\rm sg})_{\rm E1}\right),\,\nonumber\\
\ket{\Psi_{\rm Z_B}(\theta_{\rm B}, \mu_{\rm B})}_{\rm B, E2}&:=&\frac{1}{\sqrt{2}}\left(\ket{0_{\rm Z}}_{\rm B}(e^{i \theta_{\rm B}}\ket{\sqrt{\mu_{\rm B}}}_{\rm ref}\ket{e^{i \theta_{\rm B}}\sqrt{\mu_{\rm B}}}_{\rm sg})_{\rm E2}+\ket{1_{\rm Z}}_{\rm B}(\ket{e^{i \theta_{\rm B}}\sqrt{\mu_{\rm B}}}_{\rm ref}\ket{e^{i (\theta_{\rm B}+\pi)}\sqrt{\mu_{\rm B}}}_{\rm sg})_{\rm E2}\right),\,\nonumber\\
\ket{\Psi_{\rm Y_B}(\theta_{\rm B}, \mu_{\rm B})}_{\rm B, E2}&:=&\frac{1}{\sqrt{2}}\left(\ket{1_{\rm Y}}_{\rm B}(\ket{e^{i \theta_{\rm B}}\sqrt{\mu_{\rm B}}}_{\rm ref}\ket{e^{i (\theta_{\rm B}+\pi/2)}\sqrt{\mu_{\rm B}}}_{\rm sg})_{\rm E2}+\ket{0_{\rm Y}}_{\rm B}(\ket{e^{i \theta_{\rm B}}\sqrt{\mu_{\rm B}}}_{\rm ref}\ket{e^{i (\theta_{\rm B}+3\pi/2)}\sqrt{\mu_{\rm B}}}_{\rm sg})_{\rm E2}\right)\,.\nonumber
\end{eqnarray}

With these definitions, we have
\begin{eqnarray}
{\hat P}_{n_{\rm A}=0}^{({\rm E1})}\ket{\Psi_{\rm Z_A}(\theta_{\rm A}, \mu)}_{\rm A, E1}&=&\frac{e^{-\mu}}{\sqrt{2}}\left(\ket{0_{\rm Z}}_{\rm A}\ket{0,0}_{\rm E1}+\ket{1_{\rm Z}}_{\rm A}\ket{0,0}_{\rm E1}\right)=e^{-\mu}\ket{0_{\rm X}}_{\rm A}\ket{0,0}_{\rm E1}\,,
\label{appendix D eq1}\\
{\hat P}_{n_{\rm A}=0}^{({\rm E1})}\ket{\Psi_{\rm Y_A}(\theta_{\rm A}, \mu)}_{\rm A, E1}&=&\frac{e^{-\mu}}{\sqrt{2}}\left(\ket{1_{\rm Y}}_{\rm A}\ket{0,0}_{\rm E1}+\ket{0_{\rm Y}}_{\rm A}\ket{0,0}_{\rm E1}\right)=e^{-\mu}\ket{0_{\rm X}}_{\rm A}\ket{0,0}_{\rm E1}={\hat P}_{n_{\rm A}=0}^{({\rm E1})}\ket{\Psi_{\rm Z_A}(\theta_{\rm A}, \mu)}_{\rm A, E1}\,,\\
{\hat P}_{n_{\rm A}=1}^{({\rm E1})}\ket{\Psi_{\rm Z_A}(\theta_{\rm A}, \mu)}_{\rm A, E1}&=&\frac{e^{i\theta_{\rm A}}\sqrt{\mu}e^{-\mu}}{\sqrt{2}}\left[\ket{0_{\rm Z}}_{\rm A}(\ket{0,1}_{\rm E1}+\ket{1,0}_{\rm E1})-\ket{1_{\rm Z}}_{\rm A}(\ket{0,1}_{\rm E1}-\ket{1,0}_{\rm E1})\right]\nonumber\\
&=&e^{i\theta_{\rm A}}\sqrt{\mu}e^{-\mu}\left(\ket{0_{\rm Z}}_{\rm A}\ket{0_{\rm Z}}_{\rm E1}-\ket{1_{\rm Z}}_{\rm A}\ket{1_{\rm Z}}_{\rm E1}\right)/\sqrt{2}
\,,\\
{\hat P}_{n_{\rm A}=1}^{({\rm E1})}\ket{\Psi_{\rm Y_A}(\theta_{\rm A}, \mu)}_{\rm A, E1}&=&\frac{e^{i\theta_{\rm A}}\sqrt{\mu}e^{-\mu}}{\sqrt{2}}
\left[\ket{1_{\rm Y}}_{\rm A}i(\ket{0,1}_{\rm E1}-i\ket{1,0}_{\rm E1})+\ket{0_{\rm Y}}_{\rm A}(-i)(\ket{0,1}_{\rm E1}+i\ket{1,0}_{\rm E1})\right]\\
&=&ie^{i\theta_{\rm A}}\sqrt{\mu}e^{-\mu}
\left(\ket{1_{\rm Y}}_{\rm A}\ket{1_{\rm Y}}_{\rm E1}-\ket{0_{\rm Y}}_{\rm A}\ket{0_{\rm Y}}_{\rm E1}\right)
={\hat P}_{n_{\rm A}=1}^{({\rm E1})}\ket{\Psi_{\rm Z_A}(\theta_{\rm A}, \mu)}_{\rm A, E1}\,,\label{appendix D eq4}\nonumber\\
\end{eqnarray}
and the ones for systems B and E2 can be obtained with the exactly the same manner. Here, we used the identity that
$\ket{0,1}:=\ket{0_{\rm X}}$ and $\ket{1,0}:=\ket{1_{\rm X}}$.
Finally, by using these equations with the equation for ${\rm C'=0}$ in the Code mode
\begin{eqnarray}
\ket{\zeta}_{\rm C, A, E1, B, E2}&:=&\sqrt{p_{{\rm Z}_{\rm A}}p_{{\rm Z}_{\rm B}}}\ket{0_{\rm Z}}_{\rm C}\ket{\Psi_{\rm Z_A}(\theta_{\rm A}, \mu)}_{\rm A, E1}\ket{\Psi_{\rm Z_B}(\theta_{\rm B}, \mu)}_{\rm B, E2}+\sqrt{p_{{\rm Y_A}}p_{{\rm Y_B}}}\ket{1_{\rm Z}}_{\rm C}\ket{\Psi_{\rm Y_A}(\theta_{\rm A}, \mu)}_{\rm A, E1}\ket{\Psi_{\rm Y_B}(\theta_{\rm B}, \mu)}_{\rm B, E2}\,,\nonumber\\
\end{eqnarray}
we can see that the probability
of obtaining ${\rm X_C}=1$ for $(n_{\rm A}, n_{\rm B})\in \{(0,0), (1,0), (0,1), (1,1)\}$ is exactly zero. For instance, as for
$(n_{\rm A}, n_{\rm B})=(1,0)$, first note that

\begin{eqnarray}
_{\rm C}\langle 1_{\rm Z}\ket{\zeta}_{\rm A, E1, B, E2}=\sqrt{p_{{\rm Z}_{\rm A}}p_{{\rm Z}_{\rm B}}P_{{\rm Y}_{\rm A}}p_{{\rm Y}_{\rm B}}}
\ket{1_{\rm X}}_{\rm C}\left(\ket{\Psi_{\rm Z_A}(\theta_{\rm A}, \mu)}_{\rm A, E1}\ket{\Psi_{\rm Z_B}(\theta_{\rm B}, \mu)}_{\rm B, E2}-\ket{\Psi_{\rm Y_A}(\theta_{\rm A}, \mu)}_{\rm A, E1}\ket{\Psi_{\rm Y_B}(\theta_{\rm B}, \mu)}_{\rm B, E2}\right)\,.\nonumber\\
\end{eqnarray}

Then, with Eqs. (\ref{appendix D eq1})-(\ref{appendix D eq4}), we have that
\begin{eqnarray}
{\hat P}_{n_{\rm A}=1}^{({\rm E1})}{\hat P}_{n_{\rm B}=0}^{({\rm E2})} {}_{\rm C}\langle 1_{\rm Z}\ket{\zeta}_{\rm A, E1, B, E2}=0\,,
\end{eqnarray}
which concludes the proof.

\section{Explicit form of $q_{\mu_{\rm A}, \mu_{\rm B}|n_{\rm A}, n_{\rm B}}$}\label{explicit-form}
Here, we present the explicit form of $q_{\mu_{\rm A}, \mu_{\rm B}|n_{\rm A}, n_{\rm B}}$ as follows.
\begin{eqnarray}
q_{\mu_{\rm A}, \mu_{\rm B}|n_{\rm A}, n_{\rm B}}=
\frac{q_{n_{\rm A}, n_{\rm B}|\mu_{\rm A}, \mu_{\rm B}}\,\,q_{\mu_{\rm A}, \mu_{\rm B}}}{q_{n_{\rm A}, n_{\rm B}}}\,
\end{eqnarray}
with
\begin{eqnarray}
q_{n_{\rm A}, n_{\rm B}|\mu_{\rm A}, \mu_{\rm B}}&:=& e^{-2(\mu_{\rm A}+\mu_{\rm B})}\frac{(2\mu_{\rm A})^{n_{\rm A}}(2\mu_{\rm B})^{n_{\rm B}}}{n_{\rm A}! n_{\rm B}!}\,,  \label{poissonian-decoy} \\
q_{\mu_{\rm A}, \mu_{\rm B}}&:=&\frac{p_{\mu_{\rm A}} p_{\mu_{\rm B}}}{p_{\mu_{\rm A}\neq\mu_{\rm B}}+p_{\mu_{\rm A}=\mu_{\rm B}}p_{\rm T}}\, \,\,({\rm for}\,\,\mu_{\rm A}\neq\mu_{\rm B})\,, \label{prob-assign-code-test1}\\
q_{\mu_{\rm A}, \mu_{\rm B}}&:=&\frac{p_{\mu_{\rm A}} p_{\rm T}}{p_{\mu_{\rm A}\neq\mu_{\rm B}}+p_{\mu_{\rm A}=\mu_{\rm B}}p_{\rm T}}\, \,\,({\rm for}\,\,\mu_{\rm A}=\mu_{\rm B})\,,\label{prob-assign-code-test2} \\
q_{\mu_{\rm A}=\mu_{\rm B}}&:=&p_{\mu_{\rm 1}}^2+p_{\mu_{\rm 2}}^2+p_{\mu_3}^2\,,\\
q_{\mu_{\rm A}\neq\mu_{\rm B}}&:=&1-p_{\mu_{\rm A}=\mu_{\rm B}}\,,\\
q_{n_{\rm A}, n_{\rm B}}&:=&\sum_{\mu_{\rm A}, \mu_{\rm B}}^{\mu_1, \mu_2, \mu_3} q_{n_{\rm A}, n_{\rm B}|\mu_{\rm A}, \mu_{\rm B}}q_{\mu_{\rm A}, \mu_{\rm B}}\,.\label{decoy-relation-prob}
\end{eqnarray}
In Eq. (\ref{poissonian-decoy}), recall that the mean photon numbers of system E1 and E2 are defined
in terms of the double pulse and therefore we have the factor of 2 in front of the mean photon numbers.
In Eqs. (\ref{prob-assign-code-test1}) and (\ref{prob-assign-code-test2}), we take into account that
Alice and Bob perform the photon number measurements only in the Test mode within the event of
$\mu_{\rm A}=\mu_{\rm B}$.

\bibliographystyle{apsrev}

\begin{thebibliography}{99}
\bibitem{footnote} Here, one could use an alternative bound $-\log_2(1-\eta_{\rm det}e^{-\alpha L_{AB}/10})$, which includes
Charlie's detectors efficiency in the SKC bound because it affects the total loss of the channel connecting Alice to Bob via Charlie~\cite{private}. However, we adopt the conservative choice.
\end{thebibliography}

\end{document}